\documentclass[12pt]{article}
\usepackage{graphicx}
\usepackage{longtable}
\makeatletter
 
  \@addtoreset{equation}{section}
\makeatother
\def\beqar {\begin{eqnarray}}
\def\eeqar {\end{eqnarray}}
\def\beq {\begin{equation}}
\def\eeq {\end{equation}}
\def\A{{\cal A}}
\def\B{{\cal B}}
\def\C{{\cal C}}

\def\F{{\cal F}}

\def\L{{\cal L}}

\def\N{{\cal N}}

\def\S{{\cal S}}
\def\V{{\cal V}}

\def\al{\alpha}
\def\bt{\beta}
\def\del{\delta}

\def\ga{\gamma}

\def\ep{\epsilon}

\def\om{\omega}
\def\Om{\Omega}
\def\th{\theta}

\def\si{\sigma}

\def\zt{\zeta}

\def\d{\partial}

\def\Ad{{\dot A}}
\def\Bd{{\dot B}}

\def\bu{{\bar u}}

\def\hf{\frac{1}{2}}

\def\<{\langle}
\def\>{\rangle}

\def\Tr{{\rm Tr}}

\def\Path{{\rm P}}

\def\cp{{\bf CP}}

\begin{document}

\begin{titlepage}
\null\vspace{-62pt} \pagestyle{empty}
\begin{center}
\vspace{1.0truein}

{\Large\bf Holonomies of gauge fields in twistor space 5: \\
\vspace{.35cm}
\hspace{-.3cm}
amplitudes of gluons and massive scalars} \\

\vspace{1.0in} {\sc Yasuhiro Abe} \\
\vskip .12in {\it Cereja Technology Co., Ltd.\\
1-13-14 Mukai-Bldg. 3F, Sekiguchi \\
Bunkyo-ku, Tokyo 112-0014, Japan } \\
\vskip .07in {\tt abe@cereja.co.jp}\\
\vspace{1.3in}
\centerline{\large\bf Abstract}
\end{center}
Scattering amplitudes of gluons coupled with a pair of
massive scalars, so-called massive scalar amplitudes,
provide the simplest yet physically useful examples of massive amplitudes.
In this paper we construct an S-matrix functional for the massive scalar amplitudes
in a recently developed holonomy formalism in supertwistor space.
From the S-matrix functional we derive ultra helicity violating (UHV),
as well as next-to-UHV (NUHV), massive scalar amplitudes
at tree level in a form that agrees with previously known results.
We also obtain recursive expressions for non-UHV tree amplitudes in general.
These results will open up a new avenue to the study
of phenomenology in the spinor-helicity formalism.

\end{titlepage}
\pagestyle{plain} \setcounter{page}{2} 

\section{Introduction}

Recently there has been much progress in the computation
of scattering amplitudes in four-dimensional massless gauge theories
by use of the spinor-helicity formalism in twistor space.
From technical and practical perspectives,
most of the recent developments can be understood in
a form of either the CSW rules \cite{Cachazo:2004kj}
or the BCFW recursion relations \cite{Britto:2004ap,Britto:2005fq}.
In order to apply these developments to phenomenological models,
notably, in search of theories beyond the standard model of particle physics,
it is then natural to consider applications of the CSW/BCFW method
to theories with massive particles.
Indeed, such massive models were sought and investigated right after
the proposals of these methods;
for the case of the CSW rules, see \cite{Dixon:2004za,Badger:2004ty,Bern:2004ba}
and for the BCFW relations, see \cite{Badger:2005zh}-\cite{Ferrario:2006np}.
For earlier works on electroweak phenomenology in terms of the spinor-helicity
formalism, not exactly in a twistor framework, see, {\it e.g.},
\cite{Bern:1996ka,Dixon:1998py,Dittmaier:1998nn}.
Some of more recent developments along these lines
can also be found in \cite{Boels:2007pj}-\cite{Huang:2012gs}.

Of these recent investigations the simplest massive models
are presumably given by the scattering amplitudes of gluons coupled with massive scalars.
These amplitudes, which we shall call {\it massive scalar amplitudes}
from here on, are of direct relevance to one-loop calculations
in non-supersymmetric theories including QCD.
Also, the massive scalar amplitudes are closely related to multigluon amplitudes with
massive fermions, particularly quarks,
by use of the supersymmetric Ward identities \cite{Schwinn:2006ca}.
Thus a thorough and systematic understanding of
the massive scalar amplitudes is crucial to build
any phenomenological models in the spinor-helicity formalism.
Some clues to such an understanding are already known in the literature.
Particularly, Boels and Schwinn have obtained an analog of the CSW rules,
the so-called massive CSW rules, for the massive scalar amplitudes
\cite{Boels:2007pj,Boels:2008ef}.
More recently, in \cite{Kiermaier:2011cr}
Kiermaier shows that the massive CSW rules correctly
lead to the scattering amplitudes of a pair of massive
scalars and an arbitrary number of positive-helicity gluons,
the so-called {\it ultra helicity violating} (UHV) amplitudes,
whose compact expressions have been derived
previously by BCFW-type recursion methods
\cite{Badger:2005zh,Forde:2005ue,Ferrario:2006np}.
For the next-to-UHV (NUHV) massive scalar amplitudes,
their CSW-type representations are essentially obtained by
Elvang, Freedman and Kiermaier (EFK) in
the study of one-loop calculations for what is
called one-minus amplitudes in QCD \cite{Elvang:2011ub}.

Motivated by these stimulating results, in the present paper,
we consider construction of an S-matrix functional for the massive scalar amplitudes
within the framework of a recently proposed holonomy formalism in twistor space
\cite{Abe:2009kn}-\cite{Abe:2011af}.
There are a few good reasons to execute this study.
First of all, in the holonomy formalism the CSW rules are implemented
by a Wick-like contraction operator in a systematic functional language.
This implementation is not limited to tree amplitudes;
as demonstrated in \cite{Abe:2011af}, it can also be applied
to one-loop amplitudes in ${\cal N}=4$ super Yang-Mills theory.
Thus our primary concern is not the search of possible applications
of the massive CSW rules to loop amplitudes.
We would rather focus on the understanding of how the massive CSW rules
are incorporated into the holonomy formalism at tree level,
in expectation of how to obtain an insight into an utterly new
mass generation mechanism.

Secondly, a massive extension of the holonomy formalism is rather straightforward
at least from an algebraic perspective.
As in the massless case, we need to define a massive holonomy operator
so as to obtain an S-matrix functional for the massive scalar amplitudes.
Practically, this can be carried out by making a
massive extension of a bialgebraic comprehensive gauge field such that
it satisfies the infinitesimal braid relations \cite{Kohno:2002bk,Cirio:2011ja}.
As discussed in detail in section 3, it turns out that such an extension is
indeed possible, which, in turn, algebraically
guarantees the construction of the massive holonomy operator.

Lastly, we notice that our construction is in accord with
the recently studied on-shell constructibility of
massive amplitudes in general \cite{Cohen:2010mi,Boels:2011zz}.
In the holonomy formalism, physical information ({\it i.e.}, helicity and
a numbering index) is encoded in the creation operator of the involved particles.
This principle should be held even for massive particles as there are no
other ingredients for this role once a holonomy operator is defined.
This implies that we can specify the polarization of
a massive particle in a similar fashion to the case of helicity,
{\it i.e.}, we may also implement the polarization information into the
massive creation operator by
modifying Nair's prescription of superamplitudes \cite{Nair:1988bq}.
Such a modification can naturally be made by
an off-shell continuation of the null spinor momenta;
notice that one can utilize
the massive spinor-helicity formalism \cite{Dittmaier:1998nn}
to obtain an explicit form of massive spinor momenta.
We shall confirm these interpretations in section 4
by presenting an S-matrix functional for the UHV massive scalar amplitudes.

This paper is organized as follows. In section 2,
we review the foundation of the holonomy formalism.
Materials covered in this section are essential for later discussions.
In section 3, we show that the original massless
holonomy operator can naturally be extended to a massive case from
an algebraic point of view.
We then define a massive holonomy operator for gluons and massive scalars.
In section 4, we first consider off-shell continuation of Nair's
superamplitude method and
then briefly review the recent results of
the massive CSW rules by Boels and Schwinn and
their applications to the computation of
the UHV massive scalar amplitudes by Kiermaier.
We end this section by deriving an S-matrix functional for the UHV amplitudes
in terms of the above obtained massive holonomy operator.
In section 5, we extend the S-matrix functional to the NUHV amplitudes
and confirm that our computation is in accord with the EFK result.
We further discuss that the extended S-matrix also leads to recursive expressions
for non-UHV massive scalar amplitudes in general.
Lastly, we present concluding remarks.

\section{Review of holonomy formalism}

In this section, we review the foundation of the holonomy
formalism introduced in \cite{Abe:2009kn}
and developed in \cite{Abe:2009kq,Abe:2010gi,Abe:2011af}.
Materials covered here are indispensable for later discussions
but the readers who are already familiar with the holonomy
formalism may skip this reviewing section.

\noindent
\underline{Knizhnik-Zamolodchikov connections, twistor space and spinor momenta}

In the holonomy formalism, the holonomy operator refers to a
holonomy of the so-called Knizhnik-Zamolodchikov (KZ) connection
\cite{Kohno:2002bk,Cirio:2011ja}.
The KZ connection in general is defined by
\beq
    \Om =  \frac{1}{\kappa} \sum_{1 \le i < j \le n} \Om_{ij} \, \om_{ij}
    \label{2-1}
\eeq
where $\kappa$ is a non-zero constant, the so-called KZ parameter,
and $\Om_{ij}$ can be expressed as
\beq
    \Om_{ij} = a_{i}^{(+)} \otimes a_{j}^{(-)} + a_{i}^{(-)} \otimes a_{j}^{(+)}
    + 2 a_{i}^{(0)} \otimes a_{j}^{(0)} \, .
    \label{2-2}
\eeq
Here the operators $a_{i}^{( \pm )}$ and $a_{i}^{(0)}$
$(i = 1,2, \cdots , n)$ form the $SL(2, {\bf C})$ algebra:
\beq
    [ a_{i}^{(+)}, a_{j}^{(-)}] = 2 a_{i}^{(0)} \, \del_{ij}  \, , ~~~
    [ a_{i}^{(0)}, a_{j}^{(+)}] = a_{i}^{(+)} \, \del_{ij} \, , ~~~
    [ a_{i}^{(0)}, a_{j}^{(-)}] = - a_{i}^{(-)} \, \del_{ij}
    \label{2-3}
\eeq
where Kronecker's deltas show that
the non-zero commutators are obtained only when $i = j$.
The remaining commutators, those expressed otherwise, all vanish.
These operators act on a set of Fock spaces $V_i$ which are characterized by
the numbering indices $i$.
In the holonomy formalism, the operators $a_{i}^{( \pm )}$ are
identified with the creation operators of the $i$-th gluon with helicity $\pm$.
The physical Hilbert space of the holonomy formalism is then given
by $V^{\otimes n} = V_1 \otimes V_2 \otimes \cdots \otimes V_n$.
$\Om_{ij}$ in (\ref{2-2}) is a bialgebraic operator and its
action on $V^{\otimes n}$ can explicitly be written as
\beq
    \sum_{\mu} 1 \otimes \cdots \otimes 1 \otimes \rho_i (I_{\mu})
    \otimes 1 \otimes \cdots \otimes 1 \otimes \rho_j (I_{\mu}) \otimes 1 \otimes \cdots
    \otimes 1
    \label{2-4}
\eeq
where $I_\mu$ ($\mu = 0,1,2$) are elements of the $SL(2, {\bf C})$ algebra,
$\rho$ denotes its representation and $1$ denotes the identity representation.

The $\om_{ij}$'s in (\ref{2-1}) are defined by the differential one-forms:
\beq
    \om_{ij} = d \log ( z_i - z_j ) = \frac{d z_i - d z_j}{z_i - z_j}
    \label{2-5}
\eeq
where the set of complex coordinates $z_{i}$ ($i = 1,2, \cdots , n$)
are identified with local coordinates on $\cp^1$ fibers of twistor space.
In the holonomy formalism, these coordinates are related to
the homogeneous coordinates of spinor momenta for the $i$-th gluon.
The spinor momenta are parametrized in terms of null four-momenta for gluons.
One of such parametrization is given by
\beq
    u_{i}^{A} = \frac{1}{\sqrt{p_{i}^{0} - p_{i}^{3}}} \left(
            \begin{array}{c}
              {p_{i}^{1} - i p_{i}^{2}} \\
              {p_{i}^{0} - p_{i}^{3}} \\
            \end{array}
          \right)
    = \al_i \left(
            \begin{array}{c}
              1 \\
              z_i \\
            \end{array}
          \right)
    \label{2-6}
\eeq
where $A = 1,2$ and $\al_i$ is a non-zero complex number, $\al_i \in {\bf C} - \{ 0 \}$.
The null four-momentum $p_{i}^{\mu}$ $( \mu = 0, 1, 2, 3 )$ satisfies the
on-shell condition
\beq
    ( p_{i} )^{2} \, = \, \eta_{\mu \nu} p_{i}^{ \mu} p_{i}^{ \nu}
    \, = \,  ( p_{i}^{0} )^2 - ( p_{i}^{1} )^2 - ( p_{i}^{2} )^2 - ( p_{i}^{3} )^2
    \, = \, 0
    \label{2-7}
\eeq
where we use the Minkowski signature $(+---)$ for the metric $\eta_{\mu\nu}$.

Lorentz transformations of $u_{i}^{A}$ are given by
$u_{i}^{A} \rightarrow  \hat{g} u_{i}^{A}$ where $\hat{g} \in SL(2, {\bf C})$ denotes
a $(2 \times 2)$-matrix representation of $SL(2,{\bf C})$.
Scalar products of $u^A$'s, which are invariant under the $SL(2,{\bf C})$, are expressed as
\beq
    u_i \cdot u_j \equiv (u_i u_j) =   \ep_{AB} u_{i}^{A}u_{j}^{B}
    \label{2-8}
\eeq
where $\ep_{AB}$ is the rank-2 Levi-Civita tensor.
Similarly, we can define the scalar products
of the complex-conjugate spinor momenta $\bu_{i \Ad}$ $( \Ad = 1,2 )$ as
\beq
    \bu_i \cdot \bu_j \equiv [\bu_i \bu_j]  = \ep^{\Ad \Bd} \bu_{i \, \Ad}
    \bu_{j \, \Bd} \, .
    \label{2-9}
\eeq
The null four-momenta are parametrized by the combination of
the holomorphic spinor momenta $u_{i}^{A}$ and the antiholomorphic ones $\bu_{i \Ad}$.
In the spinor-helicity formalism, the four-dimensional Lorentz symmetry
is therefore given by $SL(2,{\bf C}) \times SL(2,{\bf C})$.

In terms of the holomorphic spinor momenta, the logarithmic one-forms $\om_{ij}$ in
(\ref{2-5}) can also be written as
\beq
    \om_{ij} \, = \, d \log(u_i u_j) \, = \, \frac{d(u_i u_j)}{(u_i u_j)} \, .
    \label{2-10}
\eeq

The physical configuration space of the holonomy formalism is given by
$\C^{(A)} = {\bf C}^n / \S_n $ where $n$ is the number of gauge bosons,
${\bf C}^n$ represents a set of the $z_i$ coordinates ($i= 1,2, \cdots , n$)
and $\S_n$ denotes the rank-$n$ symmetric group.
The fundamental homotopy group of $\C^{(A)}$ is given by the braid group
$\Pi_1 ( \C^{(A)} ) = \B_n$.

\noindent
\underline{Infinitesimal braid relations and the integrability of KZ connection}

The integrability of the KZ connection, {\it i.e.}, $d \Om - \Om \wedge \Om
= 0$, is guaranteed if $\Om_{ij}$ satisfies the following conditions \cite{Kohno:2002bk}:
\beqar
    \left[ \Om_{ij} , \Om_{kl} \right] &=& 0  ~~~~~ \mbox{($i,j,k,l$ are distinct),}
    \label{2-11} \\
    \left[ \Om_{ij} + \Om_{jk} , \Om_{ik} \right] &=& 0  ~~~~~ \mbox{($i,j,k$ are distinct).}
    \label{2-12}
\eeqar
These relations are known as the infinitesimal braid relations.
The commutators of bialgebraic operators are generally defined by
\beq
    [ a_i \otimes b_i , a_j \otimes b_j ]
    \, = \,
    [a_i , a_j ] \otimes b_i \otimes b_j
    + \, a_i \otimes [b_i , a_j ] \otimes b_j
    + \, a_i \otimes a_j \otimes [b_i , b_j]
    \label{2-13}
\eeq
where $a_i$ and $b_i$ $(i = 1,2,\cdots , n)$
denote a set of arbitrary operators.
From (\ref{2-2}) and (\ref{2-3}), we find that the first relation (\ref{2-11})
is obviously satisfied.
One can also check that $\Om_{ij}$'s satisfy the second relation (\ref{2-12}).

\noindent
\underline{Comprehensive gauge one-forms for gluons}

Application of these mathematical results has lead to the holonomy formalism for
gluon amplitudes.
The physical operators of gluons are given by $a_{i}^{(\pm)}$.
$\Om_{ij}$'s are not appropriate to describe gluons since
its action on the Hilbert space in (\ref{2-4})
contains the action of $a_{i}^{(0)}$.
We need to modify $\Om_{ij}$'s so that the operators $a_{i}^{(0)}$
are treated somewhat unphysically, which leads us to introduce
a ``comprehensive'' gauge one-form
\beq
    A \, = \, g \sum_{1 \le i < j \le n} A_{ij} \, \om_{ij}
    \label{2-14}
\eeq
where $g$ is a dimensionless coupling constant and $A_{ij}$ is defined as
\beq
    A_{ij} \, = \, a_{i}^{(+)} \otimes a_{j}^{(0)} + a_{i}^{(-)} \otimes a_{j}^{(0)} \, .
    \label{2-15}
\eeq

Notice that $A_{ij}$ also satisfies the infinitesimal braid relations
(\ref{2-11}), (\ref{2-12}); see \cite{Abe:2009kn} for details of its proof.
As mentioned earlier, these relations guarantee the integrability of
the ``comprehensive'' gauge field, {\it i.e.},
\beq
    DA \, =  \, dA - A \wedge A \, = \, - A \wedge A = 0
    \label{2-16}
\eeq
where $D$ denotes a covariant exterior derivative $D = d - A$.

The coupling constant $g$ is related to the KZ parameter $\kappa$ by
$g = \frac{1}{\kappa}$. For an $SU(N)$ gauge theory, this can be given by
\beq
    g \, = \, \frac{1}{\kappa} \, = \, \frac{1}{1+N} \, .
    \label{2-17}
\eeq

\noindent
\underline{Definition of the holonomy operator for $A$}

The integrability of the comprehensive gauge one-form $A$ allows us
to define a holonomy of $A$.
The holonomy operator of $A$ is defined by
\beq
    \Theta_{R, \ga}^{(A)} (u) = \Tr_{R, \ga} \, \Path \exp \left[
    \sum_{m \ge 2} \oint_{\ga} \underbrace{A \wedge A \wedge \cdots \wedge A}_{m}
    \right]
    \label{2-18}
\eeq
where $\ga$ represents a closed path on $\C^{(A)} = {\bf C}^n / \S_n$ along which the integral is
evaluated and $R$ denotes the representation of the gauge group.
The color degree of freedom
can be attached to the physical operators $a_{i}^{(\pm)}$ in (\ref{2-15}) as
\beq
    a_{i}^{(\pm)} = t^{c_i} \, a_{i}^{(\pm)c_i}
    \label{2-19}
\eeq
where $t^{c_i}$'s are the generators of the $SU(N)$ gauge group in the $R$-representation.
The symbol $\Path$ denotes an ordering of the numbering indices.
The meaning of the
action of $\Path$ on the exponent of (\ref{2-18}) can explicitly be written as
\beqar
    &&
    \Path \sum_{m \ge 2}  \oint_{\ga} \underbrace{A \wedge \cdots \wedge A}_{m}
    \nonumber \\
    &=& \sum_{m \ge 2} \oint_{\ga}  A_{1 2} A_{2 3} \cdots A_{m 1}
    \, \om_{12} \wedge \om_{23} \wedge \cdots \wedge \om_{m 1}
    \nonumber \\
    &=& \sum_{m \ge 2}  \frac{1}{2^{m+1}} \sum_{(h_1, h_2, \cdots , h_m)}
    (-1)^{h_1 + h_2 + \cdots + h_m} \,
    a_{1}^{(h_1)} \otimes a_{2}^{(h_2)} \otimes \cdots \otimes a_{m}^{(h_m)}
    \, \oint_{\ga} \om_{12} \wedge \cdots \wedge \om_{m1}
    \nonumber \\
    \label{2-20}
\eeqar
where $h_{i} = \pm = \pm 1$ ($i=1,2,\cdots, m$) denotes
the helicity of the $i$-th gluon.
In deriving the above expression, we use the relations
\beqar
    [A_{12}, A_{23}]
    &=&
    a_{1}^{(+)} \otimes a_{2}^{(+)} \otimes a_{3}^{(0)}
    - a_{1}^{(+)} \otimes a_{2}^{(-)} \otimes a_{3}^{(0)}
    \nonumber \\
    &&
    + \, a_{1}^{(-)} \otimes a_{2}^{(+)} \otimes a_{3}^{(0)}
    - a_{1}^{(-)} \otimes a_{2}^{(-)} \otimes a_{3}^{(0)} \, ,
    \label{2-21} \\
    \left[ [A_{12}, A_{23}], A_{34} \right]
    &=&
    a_{1}^{(+)} \otimes a_{2}^{(+)} \otimes a_{3}^{(+)} \otimes a_{4}^{(0)}
    - a_{1}^{(+)} \otimes a_{2}^{(+)} \otimes a_{3}^{(-)} \otimes a_{4}^{(0)}
    \nonumber \\
    &&
    - \, a_{1}^{(+)} \otimes a_{2}^{(-)} \otimes a_{3}^{(+)} \otimes a_{4}^{(0)}
    + a_{1}^{(+)} \otimes a_{2}^{(-)} \otimes a_{3}^{(-)} \otimes a_{4}^{(0)}
    \nonumber \\
    &&
    + \, a_{1}^{(-)} \otimes a_{2}^{(+)} \otimes a_{3}^{(+)} \otimes a_{4}^{(0)}
    - a_{1}^{(-)} \otimes a_{2}^{(+)} \otimes a_{3}^{(-)} \otimes a_{4}^{(0)}
    \nonumber \\
    &&
    + \, a_{1}^{(-)} \otimes a_{2}^{(-)} \otimes a_{3}^{(+)} \otimes a_{4}^{(0)}
    - a_{1}^{(-)} \otimes a_{2}^{(-)} \otimes a_{3}^{(-)} \otimes a_{4}^{(0)}
    \label{2-22}
\eeqar
and their generalization.
In the expression (\ref{2-20}), we also define
$a_{1}^{(\pm)} \otimes a_{2}^{(h_2)} \otimes
\cdots \otimes a_{m}^{(h_m)} \otimes a_{1}^{(0)}$ as
\beqar
    a_{1}^{(\pm)} \otimes a_{2}^{(h_2)} \otimes \cdots \otimes a_{m}^{(h_m)} \otimes a_{1}^{(0)}
    & \equiv &
    \hf [a_{1}^{(0)} , a_{1}^{(\pm)}] \otimes a_{2}^{(h_2)} \otimes \cdots \otimes a_{m}^{(h_m)}
    \nonumber\\
    &=&
    \pm \hf a_{1}^{(\pm)} \otimes a_{2}^{(h_2)} \otimes \cdots \otimes a_{m}^{(h_m)}
    \label{2-23}
\eeqar
where we implicitly use an antisymmetric property for the numbering indices $(1,2, \cdots , m)$.

The trace $\Tr_{R, \ga}$ in the definition (\ref{2-18}) represents
a combination of the usual color trace $\Tr_{R}$ over $t^{c_i}$'s and
the so-called braid trace $\Tr_{\ga}$ over braid generators.
The braid trace is realized by a sum over permutations of the numbering
indices; see \cite{Abe:2009kq} for details of this point.
Thus the braid trace $\Tr_\ga$ over the exponent of (\ref{2-18}) can be
expressed as
\beq
    \Tr_{\ga} \Path \sum_{m \ge 2}^{\infty} \oint_{\ga}
    \underbrace{A \wedge \cdots \wedge A}_{m}
    = \sum_{m \ge 2}
    \sum_{\si \in \S_{m-1}} \oint_{\ga}  A_{1 \si_2} A_{\si_2 \si_3} \cdots A_{\si_m 1}
    \, \om_{1 \si_2} \wedge \om_{\si_2 \si_3} \wedge \cdots \wedge \om_{\si_m 1}
    \label{2-24}
\eeq
where the summation of $\S_{m-1}$ is taken over the
permutations of the elements $\{2,3, \cdots, m \}$,
with the permutations labeled by
$\si=\left(%
\begin{array}{c}
  2 ~ 3 \, \cdots \, m \\
  \si_2 \si_3 \cdots \si_m \\
\end{array}%
\right)$.

\noindent
\underline{The holonomy operator in supertwistor space}

In the holonomy formalism,
an S-matrix functional for gluon amplitudes
is described by a holonomy operator in supertwistor space.
The supersymmetrized holonomy operator is defined by
\beq
    \Theta_{R, \ga}^{(A)} (u; x, \th ) \, = \, \Tr_{R, \ga} \, \Path \exp \left[
    \sum_{m \ge 2} \oint_{\ga} \underbrace{A \wedge A \wedge \cdots \wedge A}_{m}
    \right]
    \label{2-25}
\eeq
where the bialgebraic operator $A_{ij}$ in (\ref{2-15})
is now expressed as
\beqar
    A_{ij} &=& \sum_{\hat{h}_{i}} a_{i}^{( \hat{h}_{i} )} (x, \th) \otimes a_{j}^{(0)}
    \, ,
    \label{2-26} \\
    a_{i}^{( \hat{h}_{i} )} (x, \th) & = &
    \left. \int d \mu (p_i) ~ a_{i}^{(\hat{h}_{i})} (\xi_i) ~  e^{ i x_\mu p_{i}^{\mu} }
    \right|_{\xi_{i}^{\al} = \th_{A}^{\al} u_{i}^{A} }
    \, ,
    \label{2-27} \\
    d \mu (p_i) & \equiv &
    \frac{d^3 p_i}{(2 \pi)^3} \frac{1}{2 p_{i  0}} ~ = ~
    \frac{1}{4} \left[
    \frac{u_i \cdot d u_i}{2 \pi i} \frac{d^2 \bu_i}{(2 \pi)^2} -
    \frac{\bu_i \cdot d \bu_i }{2 \pi i} \frac{d^2 u_i }{(2 \pi)^2}
    \right]
    \, .
    \label{2-28}
\eeqar
$d \mu (p_i)$ is called the Nair measure for the null momentum $p_i$.
$a_{i}^{(\hat{h}_{i})} (x, \th)$'s are physical operators
that are defined in a four-dimensional $\N = 4$ chiral
superspace $(x, \th)$ where $x_{\Ad A}$ denote coordinates
of four-dimensional spacetime and $\th_{A}^{\al}$ $(A = 1,2; \al = 1,2,3,4)$
denote their chiral superpartners with $\N = 4$ extended supersymmetry.
These coordinates emerges from homogeneous coordinates
of the supertwistor space $\cp^{3|4}$, represented by $( u^A , v_\Ad , \xi^\al )$,
that satisfy the so-called supertwistor conditions
\beq
    v_\Ad \, = \, x_{\Ad A} u^A \, , ~~~
    \xi^\al \, = \, \th_{A}^{\al} u^A \, .
    \label{2-29}
\eeq

The physical operators $a_{i}^{(\hat{h}_{i})} (\xi_i)$ are relevant to
creations of gluons and their superpartners, having the helicity
$\hat{h}_{i} = (0, \pm \hf , \pm 1 )$. Explicitly, these supermultiplets
can be expressed as
\beqar
    \nonumber
    a_{i}^{(+)} (\xi_i) &=& a_{i}^{(+)} \, ,
    \\
    \nonumber
    a_{i}^{\left( + \frac{1}{2} \right)} (\xi_i) &=& \xi_{i}^{\al}
    \, a_{i \, \al}^{ \left( + \frac{1}{2} \right)} \, ,
    \\
    a_{i}^{(0)} (\xi_i) &=& \hf \xi_{i}^{\al} \xi_{i}^{\bt} \, a_{i \, \al \bt}^{(0)}
    \, ,
    \label{2-30}
    \\
    \nonumber
    a_{i}^{\left( - \frac{1}{2} \right)} (\xi_i) &=&
    \frac{1}{3!} \xi_{i}^{\al}\xi_{i}^{\bt}\xi_{i}^{\ga}
    \ep_{\al \bt \ga \del} \, {a_{i}^{ \left( - \frac{1}{2} \right)}}^{ \del}
    \, ,
    \\
    \nonumber
    a_{i}^{(-)} (\xi_i) &=& \xi_{i}^{1} \xi_{i}^{2} \xi_{i}^{3} \xi_{i}^{4} \, a_{i}^{(-)}
\eeqar
which are consistent with the definition of the helicity operator
\beq
    \hat{h}_{i} = 1 - \hf  u_{i}^{A} \frac{\d}{\d u_{i}^{A}} \, .
    \label{2-31}
\eeq
Use of the supermultiplets (\ref{2-30}) enables us to
define gluon operators without introducing the conventional polarization/helicity vectors.
This method is known as Nair's prescription of superamplitudes \cite{Nair:1988bq}.

\noindent
\underline{An S-matrix functional for gluon amplitudes}

In terms of the supersymmetric holonomy operator (\ref{2-25}),
an S-matrix functional for gluon amplitudes can be constructed as
\beq
    \F^{(A)} \left[  a^{(h)c}  \right]
    ~ = ~ W^{(A)} (x) \F_{MHV} \left[  a^{(h)c} \right]
    \label{2-32}
\eeq
where
\beqar
    \F_{MHV} \left[ a^{(h)c} \right]
    & = & \exp \left[ \frac{i}{g^2} \int d^4 x d^8 \th
    ~ \Theta_{R, \ga}^{(A)} (u; x ,\th) \right] \, ,
    \label{2-33}
    \\
    \widehat{W}^{(A)} (x) & = & \exp \left[-
    \int d \mu (q) \left(
    \frac{\del}{\del a_{p}^{(+)}} \otimes \frac{\del}{\del a_{-p}^{(-)}}
    \right)
    e^{- iq \cdot (x -y) }
    \right]_{y \rightarrow x}
    \nonumber  \\
    &=&
    \exp \left[-
    \int \frac{ d^4 q}{(2 \pi)^4} \frac{i}{q^2} \left(
    \frac{\del}{\del a_{p}^{(+)}} \otimes \frac{\del}{\del a_{-p}^{(-)}}
    \right)
    e^{- iq \cdot (x -y) }
    \right]_{y \rightarrow x}
    \, .
    \label{2-34}
\eeqar
Note that we take the limit $y \rightarrow x$, keeping the time ordering $x^0 > y^0$
or $x^0 - y^0 \rightarrow 0_{+}$, at the end of calculation.
The CSW rules are realized, in a functional language, by the incorporation of
the Wick-like contraction operator $\widehat{W}^{(A)} (x)$
into the S-matrix functional $\F^{(A)} \left[  a^{(h)c}  \right]$.
In (\ref{2-34}), $q$ denotes a momentum transfer which is generally off-shell
and $p$ denotes its on-shell partner. The two are related by
\beq
    q_\mu \, = \,  p_\mu + w \eta_\mu
    \label{2-35}
\eeq
where $\eta_\mu$ is a reference null-vector, satisfying $\eta^2 = 0$
and $w$ is a real number.
Since both $\eta_\mu$ and $w$
can arbitrarily be chosen, we can
fix the scaling freedom for either $\eta_\mu$ or $w$.

In terms of the S-matrix functional $\F \left[  a^{(h)c}  \right]$,
general $n$-point N$^k$MHV gluon amplitudes $(k = 0,1,2, \cdots n-4)$
are generated as
\beq
    \left. \frac{\del}{\del a_{1}^{(h_1) c_1}  } \otimes
    \frac{\del}{\del a_{2}^{(h_2) c_2} } \otimes
    \cdots \otimes \frac{\del}{\del a_{n}^{(h_n) c_n} }
    \, \F^{(A)} \left[  a^{(h)c}  \right] \right|_{a^{(h)c} =0}
    \, = \,
    \A^{(1_{h_1} 2_{h_2} \cdots n_{h_n})}_{\rm N^{k}MHV} ( x )
    \label{2-36}
\eeq
where $a^{(h)c}$ denotes a generic expression for
the gluon creation operators $a^{(h_i ) c_i}_{i}$ (with $h_i = \pm$,
$i = 1,2, \cdots , n$), which are treated as source functions in the above.
{\it Notice that the expression (\ref{2-36}) is not limited to
the case of tree amplitudes.
As shown in \cite{Abe:2011af}, the expression
is also applicable to one-loop amplitudes
and, from a functional perspective, it would and should be valid
through higher loop levels.}

In practical calculations, we need to use two key relations.
One is the normalization of the spinor momenta
\beq
    \oint_\ga d(u_1 u_2) \wedge d(u_2 u_3) \wedge \cdots \wedge d (u_m u_1) = 2^{m+1}
    \label{2-37}
\eeq
and the other is the non-vanishing Grassmann integral over $\th$'s:
\beq
    \left. \int d^8 \th  \, \xi_{r}^{1}\xi_{r}^{2}\xi_{r}^{3}\xi_{r}^{4}
    \, \xi_{s}^{1}\xi_{s}^{2}\xi_{s}^{3}\xi_{s}^{4}
    \right|_{\xi_{i}^{\al} = \th_{A}^{\al} u_{i}^{A} }
    = \, (u_r u_s )^4  \, .
    \label{2-38}
\eeq
The latter relation guarantees that gluon amplitudes
vanish unless the helicity configuration can be factorized into
the MHV helicity configurations.
Together with use of the contraction operator (\ref{2-34}),
the CSW rules are thus automatically satisfied by the Grassmann integral (\ref{2-38}).

Lastly, to clarify the notations above, we present the tree-level MHV amplitudes,
the simplest form of the gluon amplitudes,
in the $x$-space representation \cite{Abe:2009kn}:
\beqar
    \A_{\rm MHV(0)}^{(1_+ 2_+ \cdots r_{-} \cdots s_{-} \cdots n_+ )} (x)
    & \equiv &
    \A_{\rm MHV(0)}^{(r_{-} s_{-})} (x)
    \, = \,
    \prod_{i=1}^{n} \int d \mu (p_i) \, \A_{\rm MHV(0)}^{(r_{-} s_{-})} (u, \bu)
    \, ,
    \label{2-39} \\
    \A_{\rm MHV(0)}^{(r_{-} s_{-})} (u, \bu)
    & = & i g^{n-2}
    \, (2 \pi)^4 \del^{(4)} \left( \sum_{i=1}^{n} p_i \right) \,
    \widehat{A}_{\rm MHV(0)}^{(r_{-} s_{-})} (u)
    \, ,
    \label{2-40} \\
    \widehat{A}_{\rm MHV (0)}^{(r_{-} s_{-})} (u)
    & = &
    \sum_{\si \in \S_{n-1}}
    \Tr (t^{c_1} t^{c_{\si_2}} t^{c_{\si_3}} \cdots t^{c_{\si_n}}) \,
    \frac{ (u_r u_s )^4}{ (u_1 u_{\si_2})(u_{\si_2} u_{\si_3})
    \cdots (u_{\si_n} u_1)}
    \, .
    \label{2-41}
\eeqar
Of course, there exists a lot of complexity in the generalization
of these forms to non-MHV and higher-loop amplitudes but
from the above functional expression (\ref{2-36}) we can {\it in principle}
write down the $x$-space N$^k$MHV gluon amplitudes as \cite{Abe:2011af}:
\beq
    \A^{(1_{h_1} 2_{h_2} \cdots n_{h_n})}_{\rm N^{k}MHV} ( x )
    ~ = ~
    \A^{(1_{h_1} 2_{h_2} \cdots n_{h_n})}_{\rm N^{k}MHV(0)} ( x )
    \, + \,
    \A^{(1_{h_1} 2_{h_2} \cdots n_{h_n})}_{\rm N^{k}MHV(1)} ( x )
    \, + \,
    \A^{(1_{h_1} 2_{h_2} \cdots n_{h_n})}_{\rm N^{k}MHV(2)} ( x )
    \, + \, \cdots
    \label{2-42}
\eeq
where $\A^{(1_{h_1} 2_{h_2} \cdots n_{h_n})}_{\rm N^{k}MHV(L)} ( x )$
denotes the $n$-point $L$-loop N$^k$MHV gluon amplitude and
$h_i = \pm$ denotes the helicity of the $i$-th gluon,
with the total number of negative helicities being $k+2$
($k=0, 1,2, \cdots, n-4$).

These are basic results of gluon amplitudes in the holonomy formalism.
Since we consider a purely gluonic theory, the
helicity index is specified by $h_i = (+ , -)$, rather than the supersymmetric version
$\hat{h}_i = (0, \pm \frac{1}{2} , \pm )$, as shown in (\ref{2-36}) and (\ref{2-42}).
If we include massive scalars, however, we need to incorporate
$h_i = 0$ ingredients due to the definition of the helicity operator (\ref{2-31}).
Consequently, it is inevitable to modify the purely gluonic S-matrix
functional (\ref{2-32}).
In order to implement such a modification, in the next section
we first consider a massive extension of the holonomy operator.

\section{The holonomy operator of gluons and massive scalars}

In this section, we consider incorporation of massive operators
into the holonomy operator from an algebraic perspective.
The aim of this section is to construct a holonomy
operator that is relevant to
the massive scalar amplitudes, {\it i.e.},
the amplitudes of gluons coupled with massive scalar particles.

\noindent
\underline{Massive extension of comprehensive gauge fields}

To begin with, we consider a massive extension of the comprehensive
gauge field $A$ in (\ref{2-15}).
We find that the most natural extension can be made by\footnote{
The choice of $\Om$ in (\ref{2-1})
also seems reasonable at first glance since, as discussed in
the previous section, it satisfies the infinitesimal braid relations.
But calculations of $[ \Om_{12}, \Om_{23} ]$,
$\left[ [ \Om_{12}, \Om_{23}], \Om_{34} \right]$, etc.,
indicate that the holonomy operator of $\Om$ leads to unwanted
prefactors.}
\beq
    B \, = \, \sum_{1 \le i < j \le n} B_{ij} \, \om_{ij}
    \label{3-1}
\eeq
where the ``massive'' bialgebraic operator $B_{ij}$ is given by
\beq
    B_{ij} \, = \, g \left(
    a_{i}^{(+)} \otimes a_{j}^{(0)} +  a_{i}^{(-)} \otimes a_{j}^{(0)}
    \right)
    \, + \,
    a_{i}^{(0)} \otimes a_{j}^{(0)} \, .
    \label{3-2}
\eeq
As before, $\om_{ij}$ is the logarithmic one-form in (\ref{2-10}) and
$g$ denotes the dimensionless gauge coupling constant.

From the definition (\ref{3-2}), one can easily check that
$B_{ij}$ satisfy the infinitesimal braid relations:
\beqar
    \left[ B_{ij} , B_{kl} \right] &=& 0  ~~~~~ \mbox{($i,j,k,l$ are distinct),}
    \label{3-3} \\
    \left[ B_{ij} + B_{jk} , B_{ik} \right] &=& 0  ~~~~~ \mbox{($i,j,k$ are distinct).}
    \label{3-4}
\eeqar
The first relation (\ref{3-3}) is trivial from (\ref{2-3})
and (\ref{2-13}). The second part can also be checked by
\beqar
    [ B_{ij}, B_{ik} ] &=&
    g^2 [ a_{i}^{(+)} \otimes a_{j}^{(0)} ,   a_{i}^{(-)} \otimes a_{k}^{(0)} ]
    +
    g [ a_{i}^{(+)} \otimes a_{j}^{(0)} ,   a_{i}^{(0)} \otimes a_{k}^{(0)} ]
    \nonumber \\
    &&
    +
    g^2 [ a_{i}^{(-)} \otimes a_{j}^{(0)} ,   a_{i}^{(+)} \otimes a_{k}^{(0)} ]
    +
    g  [ a_{i}^{(-)} \otimes a_{j}^{(0)} ,   a_{i}^{(0)} \otimes a_{k}^{(0)} ]
    \nonumber \\
    &&
    +
    g  [ a_{i}^{(0)} \otimes a_{j}^{(0)} ,   a_{i}^{(+)} \otimes a_{k}^{(0)} ]
    +
    g  [ a_{i}^{(0)} \otimes a_{j}^{(0)} ,   a_{i}^{(-)} \otimes a_{k}^{(0)} ]
    \nonumber \\
    &=&
    0
    \label{3-5}
\eeqar
and the trivial relation $[ B_{jk}, B_{ik} ] = 0$, with
the indices $i,j,k$ being distinct.

\noindent
\underline{Definition of a holonomy operator for $B$: a first look}

Since the infinitesimal braid relations are satisfied, we can {\it naively}
define a holonomy operator of $B$ as
\beq
    \Theta_{R, \ga}^{(B)} (u) = \Tr_{R, \ga} \, \Path \exp \left[
    \sum_{r \ge 2} \oint_{\ga} \underbrace{B \wedge B \wedge \cdots \wedge B}_{r}
    \right] \, .
    \label{3-6}
\eeq

As in (\ref{2-22}) and (\ref{2-23}),
an explicit expansion of the physical operators $a_{i}^{(h_i )}$
$(h_i = \pm , 0 )$ in the integrand
can be deduced from the commutation relations
\beqar
    [ B_{12}, B_{23} ]
    &=&
    g^2 \left(
    a_{1}^{(+)} \otimes a_{2}^{(+)} \otimes a_{3}^{(0)}
    - a_{1}^{(+)} \otimes a_{2}^{(-)} \otimes a_{3}^{(0)}
    \right.
    \nonumber \\
    &&
    ~~~~~
    \left.
    + \, a_{1}^{(-)} \otimes a_{2}^{(+)} \otimes a_{3}^{(0)}
    - a_{1}^{(-)} \otimes a_{2}^{(-)} \otimes a_{3}^{(0)}
    \right)
    \nonumber \\
    &&
    + g \left(
    a_{1}^{(0)} \otimes a_{2}^{(+)} \otimes a_{3}^{(0)}
    - a_{1}^{(0)} \otimes a_{2}^{(-)} \otimes a_{3}^{(0)}
    \right) \, ,
    \label{3-7} \\
    \left[ [ B_{12}, B_{23}], B_{34} \right]
    &=&
    g^3 \left(
    a_{1}^{(+)} \otimes a_{2}^{(+)} \otimes a_{3}^{(+)} \otimes a_{4}^{(0)}
    - a_{1}^{(+)} \otimes a_{2}^{(+)} \otimes a_{3}^{(-)} \otimes a_{4}^{(0)}
    \right.
    \nonumber \\
    &&
    ~~~~~
    - \, a_{1}^{(+)} \otimes a_{2}^{(-)} \otimes a_{3}^{(+)} \otimes a_{4}^{(0)}
    + a_{1}^{(+)} \otimes a_{2}^{(-)} \otimes a_{3}^{(-)} \otimes a_{4}^{(0)}
    \nonumber \\
    &&
    ~~~~~
    + \, a_{1}^{(-)} \otimes a_{2}^{(+)} \otimes a_{3}^{(+)} \otimes a_{4}^{(0)}
    - a_{1}^{(-)} \otimes a_{2}^{(+)} \otimes a_{3}^{(-)} \otimes a_{4}^{(0)}
    \nonumber \\
    &&
    ~~~~~
    \left.
    - \, a_{1}^{(-)} \otimes a_{2}^{(-)} \otimes a_{3}^{(+)} \otimes a_{4}^{(0)}
    + a_{1}^{(-)} \otimes a_{2}^{(-)} \otimes a_{3}^{(-)} \otimes a_{4}^{(0)}
    \right)
    \nonumber \\
    &&
    + g^2 \left(
    a_{1}^{(0)} \otimes a_{2}^{(+)} \otimes a_{3}^{(+)} \otimes a_{4}^{(0)}
    - a_{1}^{(0)} \otimes a_{2}^{(+)} \otimes a_{3}^{(-)} \otimes a_{4}^{(0)}
    \right.
    \nonumber \\
    &&
    ~~~~~~~~
    \left.
    - \, a_{1}^{(0)} \otimes a_{2}^{(-)} \otimes a_{3}^{(+)} \otimes a_{4}^{(0)}
    + a_{1}^{(0)} \otimes a_{2}^{(-)} \otimes a_{3}^{(-)} \otimes a_{4}^{(0)}
    \right)
    \label{3-8}
\eeqar
and their generalization.
Terms in the leading order of $g$ are the same as the massless case.
Thus, by use of the definition (\ref{2-23}),
these terms lead to the original massless holonomy operator,
reducing the holonomy of $B$ to that of $A$.

The rest of the terms, those with $a_{1}^{(0)}$'s, would
correspond to correlators of the interaction among
gluons and a pair massive scalars.
Note that, as discussed in (\ref{2-30}) and (\ref{2-31}),
a creation operator of a scalar or spin-0 particle is described by
$a_{i}^{(0)}$ in the holonomy formalism.
If we apply the definition (\ref{2-23}), however, these
would-be massive terms vanish and we can not construct a
massive holonomy operator out of (\ref{3-6}).
This problem can be remedied by:
\begin{enumerate}
  \item considering an open path integral so that a pair of the operators
  $( a_{1}^{(0)} ,   a_{n}^{(0)})$ survives in the integrand of (\ref{3-6}); and
  \item splitting the numbering indices into those of gluons
  and massive scalars when we take a braid trace or
  a sum over the permutations of indices.
\end{enumerate}

For this purpose, we first fix the indices of massive scalars to
$1$ and $n$, being in accord with the expressions (\ref{3-7}), (\ref{3-8}).
We then identify $a_{1}^{(0)}$ and  $ a_{n}^{(0)}$
as physical operators for a pair of complex
massive scalar particles $\phi_1$ and $\bar{\phi}_n$, respectively\footnote{
We here follow the convention to use complex particles. As we shall see
later, no significant differences arise between real and complex
massive scalars in our formalism.
Use of complex scalars is simply more suitable for
the extension to amplitudes of gluons and fermions.
}.
We consider that gluons and massive scalars are
both in the $R$-representation of the gauge group,
$SU(N) \times U(1) = U(N)$, otherwise we can not properly define couplings between
them. Notice that, as in the case of scalar propagators that appears in the CSW rules,
we can assign $U(1)$ color degrees of freedom to the scalar
particles so that the single trace structure of the full amplitudes preserves.

Consequently, the braid trace in (\ref{3-6}) should be taken over the
numbering elements $\{ \si_2 , \si_3 , \cdots , \si_{r-1} , \tau_r \}
= \{ 2,3,\cdots , r \}$, satisfying the $\Path$ ordering
\beq
    \si_2 < \si_3 < \cdots < \si_{r-1} \, .
    \label{3-9}
\eeq
The braid trace can then be represented by a ``homogenous'' sum
\beq
    \sum_{ \{ \si , \tau \} }
    \, = \,
    \sum_{\tau_r = 2}^{r} \, \sum_{\si \in \S_{r-2} }
    \label{3-10}
\eeq
where $r = 3, 4, \cdots , n$.

As discussed in \cite{Abe:2010gi} (see section 3),
a product of iterated integrals over the logarithmic one-forms $\om_{ij}$'s
can be expanded, using the homogeneous sum, as
\beq
    \sum_{ \{ \si , \tau \} } \oint_\ga
    \om_{1 \si_2} \wedge \om_{\si_2  \si_3} \wedge \cdots \wedge
    \om_{\si_{r-1}  \tau_{r} } \wedge \om_{ \tau_{r} 1}
    \, = \,
    \int_{\ga_{1 r}} \om_{12} \wedge \cdots \wedge \om_{r-1 \, r}
    \int_{\ga_{r 1}} \om_{r1}
    \label{3-11}
\eeq
where $\ga_{1 r}$ and $\ga_{r 1}$
denote open paths on a physical
configuration space of interest, satisfying $\ga = \ga_{1r} \ga_{r1}$.
Notice that we split the numbering indices into $\si_i$ ($i= 2,3, \cdots , r-1$)
and $\tau_r$, respectively corresponding to the elements of gluons and
a pair of massive scalars.
In this labeling, the closed path $\ga$ can be denoted as $\ga = \ga_{\si | \tau }$.
The physical configuration is now given by
that of $(n-2)$ gluons and 2 distinct massive scalars, {\it i.e.},
\beq
    \C^{(B)} \, = \, \frac{ {\bf C}^{n-2} }{ \S_{n-2} }  \otimes {\bf C}^2
    \, = \,  {\bf C}^{n} / \S_{n-2}
    \label{3-12}
\eeq
as opposed to the pure gluonic case $\C^{(A)} = {\bf C}^{n} / \S_{n}$.
In the present massive case,
any physical observables should be symmetric under transpositions
of $(n-2)$ gluons. This is consistent with the appearance of
the sum over $\si \in \S_{r-2}$ in (\ref{3-10}).
The quantum Hilbert space, on the other hand, remains the same
as the massless case, $V^{\otimes n} = V_1 \otimes
V_2 \otimes \cdots \otimes V_n$ as discussed below (\ref{2-3}).

\noindent
\underline{Definition of a holonomy operator for $B$: a refined version}

Now that we have specified the physical configuration
on which the massive holonomy operator $\Theta_{R , \ga}^{(B)} (u)$
is defined and the quantum Hilbert space on
which $\Theta_{R , \ga}^{(B)} (u)$ acts, we
are at the stage of deriving a well-defined version of
$\Theta_{R , \ga}^{(B)} (u)$ to replace the naive guess form (\ref{3-6}).

From the above arguments, we find that
an analog of the expansion (\ref{2-24}) can be expressed as
\beqar
    &&
    \Tr_{\ga} \Path \sum_{r \ge 2}^{\infty} \oint_{\ga}
    \underbrace{B \wedge \cdots \wedge B}_{r}
    \nonumber \\
    &=&
    \sum_{r \ge 3}
    \sum_{\{ \si , \tau \}} \oint_{\ga_{\si | \tau}}  B_{1 \si_2} B_{\si_2 \si_3} \cdots
    B_{\si_{r-1} \tau_{r} } B_{\tau_{r} 1}
    ~ \om_{1 \si_2} \wedge \om_{\si_2 \si_3} \wedge \cdots \wedge
    \om_{\si_{r-1} \tau_r} \wedge \om_{\tau_r 1}
    \nonumber \\
    &=&
    \sum_{r \ge 3}
    \int_{\ga_{1 r}}  B_{1 2} B_{2 3} \cdots
    B_{r-1 \, r} ~ \om_{12} \wedge \cdots \wedge \om_{r-1 \, r}
    \int_{\ga_{r 1}} B_{\tau_{r} 1} \, \om_{r1}
    \label{3-13}
\eeqar
where we treat $B_{ij}$'s as coefficients of the logarithmic one-forms.
Since the pure gluonic part is excluded from the
physical configuration space (\ref{3-12}),
the above integral leads to operators involving gluons
coupled with a pair of massive scalars $( a_{1}^{(0)} , a_{r}^{(0)} )$.
Namely, we have
\beqar
    &&
    \Tr_{\ga} \Path \sum_{r \ge 3}^{\infty} \oint_{\ga}
    \underbrace{B \wedge \cdots \wedge B}_{r}
    \nonumber \\
    &=&
    \sum_{r \ge 3} \sum_{ ( h_2 , h_3 , \cdots , h_{r-1} ) } g^{r-2}
    \left[
    \frac{1}{2^{r-1}} ( -1 )^{ h_2 h_3 \cdots h_{r-1} }
    a_{1}^{(0)} \otimes a_{2}^{(h_2)} \otimes \cdots \otimes a_{r-1}^{(h_{r-1})}
    \otimes a_{r}^{(0)}
    \right.
    \nonumber \\
    &&
    ~ \left.  \times
    \, \int_{\ga_{1r}} \om_{12} \wedge \cdots \wedge \om_{r-1 \, r}
    \, \int_{\ga_{r1}} \om_{r1}
    \, \right]
    \nonumber \\
    &=&
    \sum_{r \ge 3} \sum_{ ( h_2 , h_3 , \cdots , h_{r-1} ) } g^{r-2}
    ( -1 )^{ h_2 h_3 \cdots h_{r-1} }
    \frac{
    a_{1}^{(0)} \otimes a_{2}^{(h_2)} \otimes \cdots \otimes a_{r-1}^{(h_{r-1})}
    \otimes a_{r}^{(0)}
    }{
    (12)(23) \cdots (r-1 \, r)( r 1)
    }
    \label{3-14}
\eeqar
where $h_i = \pm = \pm 1$ ($i=2,3, \cdots , r-1$) and we use the normalization
of the spinor momenta along an open path
\beqar
    \int_{\ga_{1r}} d(u_1 u_2) \wedge d(u_2 u_3) \wedge
    \cdots \wedge d (u_{r-1} u_{r} )
    & = &
    2^{r-1} \, ,
    \nonumber \\
    \int_{\ga_{r1}} d (u_r u_1 ) &=& 1 \, .
    \label{3-15}
\eeqar
These are open-path analogs of the closed-path normalization given in (\ref{2-37}).

To summarize, we can define the holonomy operator of gluons and massive scalars as
\beqar
    \Theta_{R , \ga}^{(B)} (u)
    &=& \exp \left[
    \sum_{r \ge 3}
    \sum_{ ( h_2 , h_3 , \cdots , h_{r-1} ) }
    g^{r-2} ( -1 )^{ h_2 h_3 \cdots h_{r-1} }
    \,
    \Tr \left(
    t^{c_2} t^{c_3} \cdots t^{c_{r-1}}
    \right)
    \right.
    \nonumber \\
    &&
    \left.
    ~~~~~~~ \times \,
    \frac{
    a_{1}^{(0)} \otimes a_{2}^{(h_2)c_2} \otimes \cdots \otimes a_{r-1}^{(h_{r-1})c_{r-1}}
    \otimes a_{r}^{(0)}
    }{
    (12)(23) \cdots (r-1 \, r)( r 1)
    }
    \right]
    \label{3-16}
\eeqar
where we make the color factor explicit.
{\it Notice that the braid trace, or a sum over permutations of gluons,
is not apparent in this form but it is already taken account of in
splitting the original closed path $\ga$ into two open paths $\ga_{1r}$ and $\ga_{r1}$.
Thus the braid trace is implicitly realized by the homogeneous sum
(\ref{3-10}) with use of the relation (\ref{3-11}).}
This form is different from the conventional color decomposition of
the massive scalar amplitudes where the sum over permutations is explicit
as in the pure gluonic case;
see, for example, \cite{Schwinn:2006ca} and references therein.

\section{An S-matrix functional for UHV tree amplitudes}

As mentioned earlier, in the holonomy formalism all the physical information
should be encoded in the creation operators, {\it i.e.}, in $a_{i}^{( h_i )}$'s
for gluons and in $( a_{1}^{(0)}, a_{n}^{(0)} )$ for massive scalars.
In the case of gluons, the helicity information is
implemented by supersymmetrization of the underlying twistor space.
As discussed in section 2, this is implemented by
Nair's superamplitude method \cite{Nair:1988bq}.
In this section, we first consider off-shell continuation of this method.
We then briefly review the recent results of
the massive CSW rules \cite{Boels:2007pj,Boels:2008ef} and
their applications to the computation of
the so-called ultra helicity violating (UHV) amplitudes, {\it i.e.},
the scattering amplitudes of a pair of massive
scalars and an arbitrary number of positive-helicity gluons, at tree level
\cite{Kiermaier:2011cr}.
To the end of this section, we shall present an S-matrix functional for the
UHV tree amplitudes by introducing a Wick-like contraction operator
involving the massive operators.

\noindent
\underline{Off-shell continuation of Nair's superamplitude method}

To begin with, we rewrite the off-shell parametrization of a four-momentum
(\ref{2-35}) as
\beq
    \widehat{p}^\mu \, = \, p^\mu + \frac{m^2}{2  ( p \cdot \eta )} \eta^\mu
    \label{4-1}
\eeq
where $\widehat{p}^\mu$ denotes a massive four-momentum with mass $m$ and
$p^\mu$ denotes its on-shell partner. $\eta^\mu$ is a reference null-vector.
In terms of spinor momenta, the null momentum is expressed as
$p^{A \Ad} = ( \si_\mu )^{A \Ad} p^\mu$, with $A$ and $\Ad$ taking
values of $(1,2)$. $\si_\mu$ here is given by $\si_\mu = ( {\bf 1} , \si_i )$
where $\si_i$ ($i=1,2,3$) and ${\bf 1}$ are the Pauli matrices and the
$(2 \times 2)$ identity matrix, respectively.
Using the parametrization (\ref{4-1}), we can then define off-shell continuation of
the null spinor momenta as \cite{Schwinn:2006ca}
\beqar
    u^A & \longrightarrow &
    \widehat{u}^A \, = \,  u^A + \frac{m}{( u \eta ) } \eta^A \, ,
    \label{4-2} \\
    \bu_\Ad & \longrightarrow &
    \widehat{\bu}^\Ad
    \, = \,  \bu^\Ad + \frac{m}{[ \bu \bar{\eta} ] } \bar{\eta}^\Ad
    \label{4-3}
\eeqar
where $\eta^A$ is a reference null spinor and $\bar{\eta}^\Ad$ is its
complex conjugate.

Since the reference null-vector $\eta^{A\Ad} = \eta^A \bar{\eta}^\Ad$
can be chosen arbitrarily, it is defined on a distinct twistor space,
decoupled from the original one that has been parametrized by
the spinors $(u^A , v_\Ad)$ satisfying the condition $v_\Ad = x_{\Ad A} u^A$.
This interpretation of $\eta^{A \Ad}$ is in accord with
the definitions (\ref{4-1})-(\ref{4-3}).
In order to construct a massive model in the spinor-helicity formalism,
however, naive substitution of $u^A$'s by $\widehat{u}^A$'s does not work out well.
For example, one can consider that an off-shell continuation of the
the projected Grassmann variable $\xi^\al = \th^\al_A u^A$ in (\ref{2-29}) is
given by $\widehat{\xi}^\al  =   \th^\al_A \widehat{u}^A$.
Use of $\widehat{\xi}^\al$ in the expressions of Nair's
superamplitude method (\ref{2-30}) leads to vanishing
UHV amplitudes due to the Grassmann integral (\ref{2-38}).
But this is contradictory because, as reviewed below,
the UHV amplitudes are non-vanishing in general.

Simple use of (\ref{4-2}) and (\ref{4-3}) therefore does not
lead to massive extensions in the spinor-formalism.
In fact, one should rather think of two distinct sets of
twistor variables $( u^A , v_\Ad )$ and $( w^A , \pi_\Ad )$
where $w^A = \frac{m}{(u \eta )} \eta^A$ and $\pi_\Ad = x_{\Ad A} w^A$.
In other words, we should use a two-spinor basis spanned by \cite{Boels:2011zz}
\beq
    \left\{ u^A  \, , ~ \frac{m}{( u \eta ) } \eta^A   \right\} \, , ~~~
    \left\{ \bu^\Ad \, , ~ \frac{m}{[ \bu \bar{\eta} ] } \bar{\eta}^\Ad \right\}
    \label{4-4}
\eeq
to describe holomorphic and antiholomorphic massive quantities,
respectively.

Notice that the four-dimensional spacetime $x_{A\Ad}$ emerges
from each of the twistor variables. This feature should be preserved
after supersymmetrization of the underlying twistor spaces.
Namely, in addition to the original supertwistor variables
$( u^A , v_\Ad , \xi^\al )$, we need to
introduce new supertwistor variables $( w^A , \pi_\Ad , \zt^\al )$ such that
the supertwistor conditions
\beq
    \pi_\Ad \, = \, x_{\Ad A} w^A \, = \, x_{\Ad A} \frac{m}{( u \eta ) } \eta^A
    \, , ~~~
    \zt^\al \, = \, \th_{A}^{\al} w^A
    \, = \, \th_{A}^{\al} \frac{m}{( u \eta ) } \eta^A
    \label{4-5}
\eeq
are satisfied ($\al = 1,2,3,4)$.
The emergent chiral superspace, {\it i.e.},
the four-dimensional spacetime $x_{A\Ad}$ and its chiral superpartner $\th_A^\al$,
is identical for either the original or the new supertwistor spaces.
This is explicitly presented in (\ref{2-29}) and (\ref{4-5}).

We now consider off-shell continuation of Nair's superamplitude method.
Based on the above arguments, this can be implemented by modifying
the operators (\ref{2-30}) for massive scalars in terms of $\xi^\al$'s and $\zt^\al$'s.
For this purpose, we take account of the conditions that
(a) the UHV tree amplitudes are non-vanishing
and (b) the massive scalar operators have
2 degrees of homogeneity in $u$'s.
The latter condition is in accord with the helicity operator (\ref{2-31}).
Regarding the former, we shall show an explicit form of the UHV tree amplitudes later;
see (\ref{4-28}).
Using the Grassmann integral (\ref{2-38}), we then find that
the massive scalar operators can uniquely be determined as
$a_{i \, \al \bt}^{(0)} ( \xi_i , \zt_i ) = \hf \xi_i^\al \xi_i^\bt \, a_{i \, \al \bt}^{(0)}$
where $a_{i \, \al \bt}^{(0)} = \frac{1}{12} \ep_{\al \bt \ga \del} \xi_i^\ga
\zt_i^\del \, a_{i}^{(0)}$. In other words, we can define an off-shell
continuation of the operator $a_{i}^{(0)} ( \xi_i )$ in (\ref{2-30}) as
\beq
    a_{i}^{(0)} ( \xi_i , \zt_i ) ~ = ~
    \xi_{i}^{1} \xi_{i}^{2} \xi_{i}^{3} \zt_{i}^{4} \, a_{i}^{(0)}
    \label{4-6}
\eeq
where we shall specify the numbering index to $i = 1, n$ for massive scalars.

On the other hand, the gluon operators remain the same as in (\ref{2-30}), {\it i.e.},
\beqar
    a_{i}^{(+)} (\xi_i) & = & a_{i}^{(+)} \, ,
    \label{4-7}\\
    a_{i}^{(-)} (\xi_i) & = & \xi_{i}^{1} \xi_{i}^{2} \xi_{i}^{3} \xi_{i}^{4} \, a_{i}^{(-)}
    \label{4-8}
\eeqar
where $i = 2,3, \cdots , n-1$ (with $n= 3,4, \cdots$).
The gluonic part of the massive holonomy operator (\ref{3-16}) is
then automatically obtained by use of the $\N = 4$ chiral superspace
representation (\ref{2-27}) with the on-shell Nair measure (\ref{2-28}).
For the massive scalars, the same superspace representation can be obtained by
use of off-shell continuation of the Nair measure $d \mu ( \widehat{p}_i )$.
(Although we shall not use the off-shell Nair measure explicitly in the
present paper, interested reader may refer to details of
the off-shell Nair measure in \cite{Abe:2011af}.)
The chiral superspace representation of the massive operators
can then be expressed as
\beqar
    a_{i}^{( 0 )} (x, \th)  & = &
    \left. \int d \mu ( \widehat{p}_i ) ~ a_{i}^{(0)} ( \xi_i , \zt_i )
    ~  e^{ i x_\mu \widehat{p}_i^\mu }
    \right|_{ \xi_{i}^{\al} = \th_{A}^{\al} u_i^A ,
    \, \zt_{i}^{\al} = \th_{A}^{\al} w_i^A  }
    \, ,
    \label{4-9} \\
    \widehat{p}_i^\mu & = & p_{i}^{\mu} + \frac{m^2}{2 (p_i \cdot \eta_i )} \eta_i^\mu
    \, ,
    \label{4-10} \\
    w_i^A & = &   \frac{m}{( u_i \eta_i ) } \eta_i^A
    \, .
    \label{4-11}
\eeqar

{\it
We can use the expressions (\ref{2-27}) and (\ref{4-9})
for gluons and massive scalars, respectively, to
construct a supersymmetric version of the massive holonomy operator.
Namely, we can obtain
the supersymmetric massive holonomy operator $\Theta_{R , \ga}^{(B)} (u; x, \th)$
out of $\Theta_{R , \ga}^{(B)} (u)$ in (\ref{3-16})
with replacements of $\{ a_{i}^{(\pm )} , a_{j}^{( 0 )} \}$
by $\{ a_{i}^{(\pm )}  ( x, \th ) , a_{j}^{(0)}  ( x, \th ) \}$
where $i=2,3,\cdots , r-1$ and $j = 1, r$.
}

\noindent
\underline{Review of the massive CSW rules}

In the following, we briefly review the massive CSW rules
of Boels and Schwinn \cite{Boels:2007pj,Boels:2008ef}.
These are an analog of the original CSW rules
for gluons $g_i^{\pm}$ and massive complex scalars $\phi_i$, ${\bar \phi}_i$.
As in the original case, the massive CSW rules give prescription
for amplitudes in terms of vertices connected by
massless and massive scalar propagators,
\beq
    D_{g^+ g^-} ( \widehat{p}^2 ) \, = \, \frac{i}{\widehat{p}^2} \, , ~~~~~
    D_{{\bar \phi} \phi} (\widehat{p}^2 ) \, = \, \frac{i}{ \widehat{p}^2 - m^2}
    \label{4-12}
\eeq
for positive and negative-helicity gluons and a pair of
massive scalars, respectively.
Up to constant factors, the involving vertices are expressed as
\beqar
    V_{\rm MHV}
    ( g_{1}^{+} g_{2}^{+} \cdots g_{i+1}^{+} g_{i}^{-}
    g_{i+1}^{+} \cdots g_{j-1}^{+} g_{j}^{-} g_{j+1}^{+} \cdots  g_{n}^{+} )
    &=&
    \frac{ (ij)^4 }{(12)(23) \cdots (n1)} \, ,
    \label{4-13} \\
    V_{\rm UHV}
    ( \bar{\phi}_{1} g_{2}^{+} \cdots g_{n-1}^{+}  \phi_{n} )
    &=&
    m^2 \frac{ (n1)^2 }{(12)(23) \cdots (n1)}
    \label{4-14}
\eeqar
where $V_{\rm MHV}$ is a purely gluonic MHV vertices, with its form
being the same as the original CSW rules.
Peculiarity of the massive CSW rules lies in the form of
$V_{\rm UHV}$ which is proportional to $m^2$.
There also exist non-UHV type vertices involving the massive CSW rules
\beqar
    &&
    V_{\rm NUHV}^{\rm (BS)}
    ( \bar{\phi}_{1} g_{2}^{+} \cdots g_{i+1}^{+} g_{i}^{-}
    g_{i+1}^{+} \cdots  g_{n-1}^{+}  \phi_{n} )
    \nonumber \\
    && ~~~ = ~
    - \frac{ (1i)^2 (in)^2 }{(12)(23) \cdots (n1)} \, ,
    \label{4-15}
    \\
    &&
    V_{\rm UHV^2}^{\rm (BS)}
    ( \bar{\phi}_{1} g_{2}^{+} \cdots g_{i+1}^{+} \phi_{i}
    \bar{\phi}_{i+1} g_{i+2}^{+} \cdots  g_{n-1}^{+}  \phi_{n} )
    \nonumber \\
    && ~~~ = ~
    - \hf \left(
    \frac{ (1i)^2 (i+1 \, n)^2 }{(12)(23) \cdots (n1)}
    +
    \frac{ (1i) (i+1 \, n) (1 \, i+1 ) (i n) }{(12)(23) \cdots (n1)}
    \right)
    \label{4-16}
\eeqar
which are not of direct relevance to the calculations of
the UHV massive scalar amplitudes.
In the above expressions the spinor momenta corresponding to the massive scalars,
{\it i.e.}, $u^A_1$ and $u^A_n$ are given by the on-shell partners of
the actual massive spinor momenta
$\widehat{u}^A_1$ and $\widehat{u}^A_n$, respectively.
Notice that these spinor momenta are related to each other by (\ref{4-2}).

The massive CSW rules have been proposed by use of two Lagrangian-based methods.
One is to use a canonical transformation in the light-cone gauge
\cite{Gorsky:2005sf,Mansfield:2005yd,Ettle:2006bw} and
the other is to use an action constructed in twistor space
\cite{Mason:2005zm,Boels:2007qn}.
In either approach, one starts from the ordinary
Lagrangian for gluons and massive scalars
\beq
    \L  \, = \,
    - \frac{1}{4 g^2} F_{\mu \nu}^{c} F^{c \, \mu \nu} +
    ( \overline{ D_\mu \phi } )^c (D^\nu \phi )^c  - m^2 \bar{\phi}^c \phi^c
    \label{4-17}
\eeq
where $D_\mu = \d_\mu + A_\mu$ ($A_\mu = - i t^c A_\mu^c$) is the covariant
derivative and $F_{\mu\nu} = [ D_\mu , D_\nu ] = -i t^c F_{\mu \nu}^{c}$
is the field strength for gluons. Following the notation (\ref{2-19}), we here
denote the color factor by $t^c$'s.
One then carries out field redefinitions
of $\bar{\phi}$ and $\phi$ such that non-MHV type couplings between a pair of
massive scalar fields and gluons are eliminated. Together with
supersymmetry arguments, this enables one to obtain
the above forms of vertices \cite{Boels:2008ef}.

Since the massive CSW rules are based on the Lagrangian formalism,
they are not necessarily compatible with our holonomy formalism
where we do/can not introduce massive potentials.
In fact, using our parametrization (\ref{4-6}) and the Grassmann
integral (\ref{2-38}), we can readily find that the vertices
(\ref{4-15}) and (\ref{4-16}) vanish, {\it i.e.},
\beqar
    V_{\rm NUHV} ( \bar{\phi}_{1} g_{2}^{+} \cdots g_{i+1}^{+} g_{i}^{-}
    g_{i+1}^{+} \cdots  g_{n-1}^{+}  \phi_{n} )
    & = & 0 \, ,
    \label{4-18} \\
    V_{\rm UHV^2} ( \bar{\phi}_{1} g_{2}^{+} \cdots g_{i+1}^{+} \phi_{i}
    \bar{\phi}_{i+1} g_{i+2}^{+} \cdots  g_{n-1}^{+}  \phi_{n} )
    & = & 0
    \label{4-19}
\eeqar
where we omit the suffix ${\rm (BS)}$ to distinguish the vertices
from those of the massive CSW rules.
These results seem to contradict each other.
In fact, however, although it is not well-recognized in the literature,
there are no explicit derivations of the non-UHV type
vertices $V_{\rm NUHV}^{\rm (BS)}$ and $V_{\rm UHV^2}^{\rm (BS)}$,
as clearly stated in \cite{Boels:2008ef} (see at the end of subsection 3.3).
The NUHV vertex $V_{\rm NUHV}^{\rm (BS)}$, together with the UHV vertex
$V_{\rm UHV}$, does lead to
four- and five-point NUHV massive scalar amplitudes \cite{Boels:2007pj,Boels:2008ef}
but there are also possibilities that different
NUHV vertices would lead to correct NUHV tree amplitudes
because these amplitudes are generally dependent upon reference spinors
which we can arbitrarily choose.
We shall come back this point and consider such possibilities in the next section;
see discussions below (\ref{5-8}).

As far as the UHV amplitudes are concerned, this apparent discrepancy goes away.
To see this assertion, we now present a functional derivation of the UHV vertex
(\ref{4-14}) in terms of the massive holonomy operator (\ref{3-16}).

\noindent
\underline{Choice of reference spinors and functional derivation of the UHV vertex}

We first fix the reference null-vector corresponding to the pair of massive
scalar particles by
\beq
    \eta_1^\mu = p_n \, , ~~~~ \eta_n^\mu = p_1 \ .
    \label{4-20}
\eeq
This means that we have
\beqar
    \widehat{p}_1^\mu &=& p_1^\mu + \frac{m^2}{2 (p_1 \cdot p_n )} p_n^\mu
    \, = \, p_1^\mu + w p_n^\mu  \, ,
    \label{4-21} \\
    \widehat{p}_n^\mu &=& p_n^\mu + \frac{m^2}{2 (p_n \cdot p_1 )} p_1^\mu
    \, = \, p_n^\mu  + w p_1^\mu  \, ,
    \label{4-22}
\eeqar
satisfying $\widehat{p}_1^2 = \widehat{p}_n^2 = m^2$ and $m = \frac{m^2}{2 (p_1 \cdot p_n )}$.
Since both $\widehat{p}_1^\mu$ and $\widehat{p}_n^\mu$ are massive, we can parametrized
them as (\ref{4-21}) and (\ref{4-22}) in a suitable reference frame.
This parametrization is qualitatively different from
off-shell prescription for virtual gluons where we set all reference null-vectors
identical.

Fixing the reference spinors as such, we now derive the UHV vertex (\ref{4-14})
from the supersymmetric version of
the massive holonomy operator (\ref{3-16}) by a functional method.
As in the MHV amplitudes, we introduce a generating functional
\beq
    \F_{\rm UHV}^{\rm (vertex)} \left[ a^{( \pm )c} , a^{(0)} \right]
    \, = \,
    \exp \left[
    i \int d^4 x d^8 \th \, \Theta_{R , \ga}^{(B)} (u; x, \th)
    \right] \, .
    \label{4-23}
\eeq
Then the UHV vertex can be generated as
\beqar
    &&
    \left.
    \frac{\del}{\del a_{1}^{(0)}} \otimes
    \frac{\del}{\del a_{2}^{(+)}} \otimes
    \frac{\del}{\del a_{3}^{(+)}} \otimes \cdots \otimes
    \frac{\del}{\del a_{n-1}^{(+)}} \otimes
    \frac{\del}{\del a_{n}^{(0)}}
    \, \F_{\rm UHV}^{\rm (vertex)} \left[ a^{( \pm )c} , a^{(0)} \right]
    \right|_{a^{( \pm )c} = a^{(0)} = 0}
    \nonumber \\
    & \equiv &
    \V_{\rm UHV}^{( \bar{\phi}_{1} g_{2}^{+} \cdots g_{n-1}^{+}  \phi_{n} )} (x)
    ~ = ~
    \int d \mu (\widehat{p}_1 )
    \prod_{i=2}^{n-1}  d \mu ( p_i )
    d \mu (\widehat{p}_n )
    \, \V_{\rm UHV}^{(  \bar{\phi}_1 \phi_n )} ( u , \bu ) \, ,
    \label{4-24}
\eeqar
\beqar
    \V_{\rm UHV}^{( \bar{\phi}_1 \phi_n  )} ( u , \bu )
    &=&
    - i g^{n-2}
    \, (2 \pi)^4 \del^{(4)} \left( \widehat{p}_1 +
    \sum_{i=2}^{n-1} p_i + \widehat{p}_n \right) \,
    \widehat{V}_{\rm UHV}^{( \bar{\phi}_1  \phi_n )} (u)
    \, ,
    \label{4-25} \\
    \widehat{V}_{\rm UHV}^{( \bar{\phi}_1 \phi_n )} (u)
    & = &
    \Tr ( t^{c_{2}} t^{c_{3}} \cdots t^{c_{n-1}}) \,
    \frac{ m^2 \, (n 1)^2}{ (12)(23) \cdots (n-1 \, n) (n 1) }
    \nonumber \\
    & = &
    \Tr ( t^{c_{2}} t^{c_{3}} \cdots t^{c_{n-1}}) \,
    V_{\rm UHV}
    ( \bar{\phi}_{1} g_{2}^{+} \cdots g_{n-1}^{+}  \phi_{n} )
    \label{4-26}
\eeqar
where we use the Grassmann integral
\beqar
    &&
    \int d^8 \th \, \xi^1_1 \xi^2_1 \xi^3_1 \zt^4_1
    \, \xi^1_n \xi^2_n \xi^3_n \zt^4_n \,
    \nonumber \\
    &=&
    (1 n)^3 \int d^2 \th \, \frac{m}{(u_1 \eta_1)} \eta_1^A \th_A \,
    \frac{m}{(u_n \eta_n )} \eta_n^B \th_B
    \nonumber \\
    &=&
    (1 n)^3 \frac{m^2}{( u_1 u_n ) ( u_n u_1 )} \underbrace{
    \int d^2 \th \, u_n^A \th_A \, u_1^B \th_B
    }_{= \, ( u_n u_1)}
    ~ = ~
    m^2 \, (n 1)^2 \, .
    \label{4-27}
\eeqar
This Grassmann integral guarantees that only the UHV-type vertices
survive upon the evaluation of functional derivatives in (\ref{4-24}).
This also automatically leads to the vanishing of non-UHV vertices
(\ref{4-18}) and (\ref{4-19}), rather than (\ref{4-15}) and (\ref{4-16}).

Another interesting feature in the expressions (\ref{4-24})-(\ref{4-27})
is that there arise no sums over permutations of the numbering indices,
contrary to the case of MHV gluon amplitudes (\ref{2-39})-(\ref{2-41}).
This implies that the number of terms to describe the UHV
massive scalar amplitudes drastically decreases from that of
the MHV gluon amplitudes.
However, such a reduction does not occur in the massive scalar amplitudes.
This is due to the fact that we can construct the UHV amplitudes
by connecting the UHV vertices with as-many-as-possible
massive propagators $D_{\bar{\phi} \phi} (\widehat{p})$ in (\ref{4-12}).
Notice that the number of propagators or vertices is independent of
the gluon helicity configurations in the present case, while
in the gluon amplitudes the number of massless propagators
$D_{g^+ g^-} ( \widehat{p}^2 )$ in (\ref{4-12})
is fixed by the helicity configurations
or by the number of negative-helicity gluons.

We can then express the UHV amplitudes by a UHV vertex expansion.
As we shall review in a moment, indeed, such an expansion is explicitly realized in
Kiermaier's expression for the UHV amplitudes \cite{Kiermaier:2011cr}.
Once the UHV amplitudes are constructed in this way, extension
to next-to-UHV (NUHV) amplitudes which contains one negative-helicity gluon
in addition to the UHV configuration, is straightforward by application
of the original CSW rules or the MHV rules to the gluonic part of the amplitudes.
Generalization to N$^k$UHV amplitudes ($k = 1,2, \cdots , n-3$) can be
carried out in the same manner.
We can therefore construct the scattering amplitudes
of an arbitrary number of gluons in any helicity configurations and
a pair of complex massive scalars.
We may call this construction the ``UHV rules'' for massive scalar amplitudes
in analogy to the MHV rules for gluon amplitudes.
We shall consider the non-UHV type constructions in the next section.

\noindent
\underline{The UHV expansion: Kiermaier's result for the UHV tree amplitudes}

Recently Kiermaier shows that the UHV tree amplitudes can be
obtained by use of the massive CSW rules, more precisely,
by use of the UHV vertex (\ref{4-14}) and the massive propagator (\ref{4-12}).
The resultant expression is given by \cite{Kiermaier:2011cr}
\beq
    A_{{\rm UHV} (0)}^{( \bar{\phi}_1 g_2^+ \cdots g_{n-1}^{+} \phi_{n} ) }
    \, = \,
    \frac{ - m^2 }{(12)(23)\cdots (n-1 \, n)}
    \left( 1 \left|
    \prod_{j = 2}^{n-2} \left[
    1 - \frac{  i m^2 |J ) ( j \, j+1 ) (J | }
    {(\widehat{P}_J^2 - m^2) (j J) (J \, j+1 ) }
    \right]
    \right| n \right)
    \label{4-28}
\eeq
where $(1 | n ) = (1 n) = \ep_{AB} u_1^A u_n^B$ and
\beq
    \widehat{P_J}^\mu \, \equiv \, \widehat{p}_1^\mu + p_2^\mu + p_3^\mu
    + \cdots + p_j^\mu
    \, .
    \label{4-29}
\eeq
As before, the on-shell partner of $\widehat{P_J}$ is defined as
\beq
    \widehat{P_J}^\mu \, = \, p_J^\mu + \frac{m^2}{2 ( p_J \cdot \eta_J ) } \eta_J^\mu
    \label{4-30}
\eeq
where $\eta_J^\mu$ denotes a reference null-vector.
The corresponding spinor momenta $(u_J^A , \bu_J^\Ad )$ are then defined by
\beq
    p_{J}^{A \Ad} \, = \, u_J^A \bu_J^\Ad
    \, .
    \label{4-31}
\eeq

While the form (\ref{4-28}) is probably the most concise expression
of the UHV tree amplitude, for the clarification of the above mentioned
``UHV rules,'' we now rewrite it as follows:
\beq
    A_{{\rm UHV} (0)}^{ ( \bar{\phi}_1 g_2^+ \cdots g_{n-1}^{+} \phi_{n} ) }
    \, = \,
    \frac{ m^2 (n1) }{(12)(23)\cdots (n1)}
    \widehat{(n1)}
    \label{4-32}
\eeq
where
\beqar
    \widehat{(n1)}
    &=&
    (n1) \, + \, \sum_{j =2}^{n-2}
    \frac{(J1)}{(jJ)}
    \frac{i m^2 (j \, j+1)}{ \widehat{P}_J^2 - m^2}
    \frac{(nJ)}{(J\, j+1)}
    \nonumber \\
    &&
    + \,
    \sum_{2 \le i < j \le n-1 }
    \frac{(I1)}{(iI)}
    \frac{ i m^2 (i \, i+1)}{ \widehat{P}_I^2 - m^2}
    \frac{(JI)}{(I \, i+1)(j J)}
    \frac{ i m^2 (j \, j+1)}{ \widehat{P}_J^2 - m^2}
    \frac{(nJ)}{(J \, j+1)}
    \nonumber \\
    &&
    + \, \sum_{2 \le i < j < k \le n-1 }  \, \Biggl[ \,
    \frac{(I1)}{(iI)}
    \frac{ i m^2 (i \, i+1)}{ \widehat{P}_I^2 - m^2}
    \frac{(JI)}{(I \, i+1)(j J)}
    \frac{ i m^2 (j \, j+1)}{ \widehat{P}_J^2 - m^2}
    \nonumber \\
    &&
    ~~~~~~~~~~~~~~~~~~~~~~~~~~~
    \times \,
    \frac{(KJ)}{(J \, j+1)(k K)}
    \frac{ i m^2 (k \, k+1)}{ \widehat{P}_K^2 - m^2}
    \frac{(nK)}{(K \, k+1)}
    \,   \Biggr]
    \nonumber \\
    &&
    + \, \cdots \, .
    \label{4-33}
\eeqar
Note that the factor $i m^2$ in the numerators should
be regarded as $i m^2 = \sqrt{-1} m^2$. The uppercase letters
$I,J,K, \cdots$ play the same role as the $J$ in (\ref{4-29})-(\ref{4-31}).
More explicitly we can expand the UHV tree amplitudes as
\beqar
    &&
    A_{{\rm UHV} (0)}^{ ( \bar{\phi}_1 g_2^+ \cdots g_{n-1}^{+} \phi_{n} ) }
    \nonumber \\
    &=&
    ~
    \frac{ m^2 (n1)^2 }{(12)(23)\cdots (n1)}
    \nonumber \\
    &&
    +
    \sum_{j = 2}^{n-2}
    \frac{ m^2 (J1)^2 }{(12)(23)\cdots (j \, j-1) (j J)(J1)}
    \frac{ i }{ \widehat{P}_J^2 - m^2}
    \frac{ m^2 (nJ)^2 }{(J \, j+1)(j+1 \, j+2) \cdots (n-1 \, n) (n J)}
    \nonumber \\
    &&
    +
    \sum_{2 \le i < j \le n-1 } \, \Biggl[
    \frac{ m^2 (I1)^2 }{(12) \cdots  (i I)(I1)}
    \frac{ i }{ \widehat{P}_I^2 - m^2}
    \frac{ m^2 (JI)^2 }{(I \, i+1) (i+1 \, i+2) \cdots (j-1 \, j)(jJ) (JI)}
    \nonumber \\
    &&
    ~~~~~~~~~~~~~~~~~~~~~~~~~~~~~~~~~
    \times \,
    \frac{ i }{ \widehat{P}_J^2 - m^2}
    \frac{ m^2 (nJ)^2 }{(J \, j+1)(j+1 \, j+2) \cdots (n-1 \, n) (nJ)}
    \,   \Biggr]
    \nonumber \\
    &&
    +
    \sum_{2 \le i < j < k \le n-1 } \, \Biggl[
    \frac{ m^2 (I1)^2 }{(12) \cdots  (i I)(I1)}
    \frac{ i }{ \widehat{P}_I^2 - m^2}
    \frac{ m^2 (JI)^2 }{(I \, i+1) (i+1 \, i+2) \cdots (j-1 \, j)(jJ) (JI)}
    \nonumber \\
    &&
    ~~~~~~~~~~~~~~~~~~~~~~~~~~~~~~~~~
    \times \,
    \frac{ i }{ \widehat{P}_J^2 - m^2}
    \frac{ m^2 (KJ)^2 }{(J \, j+1) (j+1 \, j+2) \cdots (k-1 \, k)(k K) (KJ)}
    \nonumber \\
    &&
    ~~~~~~~~~~~~~~~~~~~~~~~~~~~~~~~~~
    \times \,
    \frac{ i }{ \widehat{P}_K^2 - m^2}
    \frac{ m^2 (nK)^2 }{(K \, k+1)(k+1 \, k+2) \cdots (n-1 \, n) (nK)}
    \,   \Biggr]
    \nonumber \\
    &&
    + \cdots \, .
    \label{4-34}
\eeqar
One can straightforwardly obtain the higher terms,
those terms higher in the number of propagators or
UHV vertices. The total number of terms involved in the expansion
(\ref{4-34}) can be calculated as
\beq
    \sum_{k = 0}^{n-3} {}_{n-3}C_k \, = \, 2^{n-3}
    \label{4-35}
\eeq
where ${}_{n-3}C_k$ denotes the number of $k$-combinations out of
$(n-3)$ elements which is also denoted as $C(n-3, k)$.
As expected, this is equivalent to that of the expression (\ref{4-28})
since we can easily count it as $(1+1)^{n-3}$.

As mentioned in section 3, there are no apparent sums over permutations of
number indices, or braid traces, in the definition of the massive holonomy operator.
Such sums are already taken account of in the product of iterated integrals (\ref{3-11}).
The relevant sum is given by the homogeneous sum (\ref{3-10})
which, if explicit, produces $(n-1)!$ terms for the $n$-point UHV tree amplitude.
In fact, this is what happens in the MHV tree amplitudes of gluons as well
since the $n$-point MHV amplitude has
$(n-1)!$ terms due to its braid trace or the sum over $\si \in \S_{n-1}$.

The UHV and the MHV amplitudes are different in structure,
the physical configuration spaces are respectively given by
$\C^{(B)} = {\bf C}^n / \S_{n-2}$ and $\C^{(A)} = {\bf C}^n / \S_n$, respectively.
Yet it is interesting to see that the above factor of $2^{n-3}$ arises from
the absorption of the braid trace in a sort of compensating manner.
It is also intriguing to compare the numbers involving
terms for the UHV and the MHV amplitudes. These are given by
by $2^{n-3}$ and $(n-1)!$, respectively.
The logarithm of these can be
evaluated as $(n-3) \ln 2 < (n-1) [ \ln (n-1) - 1 ]$ for a large $n$.

From the expansion (\ref{4-34}) one can easily visualize the
expansion or the clusterization
of the UHV tree amplitudes in terms of the UHV vertices
connected by the massive propagators.
This expansion is exactly what has been found in
the derivation of Kiermaier's expression (\ref{4-28})
in comparison with previously known results for
the UHV tree amplitudes \cite{Badger:2005zh,Forde:2005ue,Ferrario:2006np}.
In the following, we interpret this UHV expansion in a functional language.

\noindent
\underline{Contraction of massive scalar operators}

As in the CSW rules for gluons, we can and should introduce
a contraction operator involving the massive scalar operators.
We can, for example, contract a pair of $a^{(0)}_{J}$
and $a^{(0)}_{-J}$ ($2 \le J \le n-2$) to
replace it by a massive scalar propagator,
with its momentum transfer given by $\widehat{p}_J^\mu$ in
(\ref{4-29}) or (\ref{4-30}).
Notice that once $J$ is chosen, we consider the numbering indices
in modulo $J$, {\it i.e.},
\beq
    J \, \equiv \, -J \, \equiv \, 0 ~~~ (mod ~ J)
    \label{4-36}
\eeq
and generally $J + i \equiv - J + i \equiv i$ ($mod ~ J$) for
$0 \le i \le J$.

In analogy to the CSW rules (\ref{2-34}),
such a contraction operator can be defined as
\beq
    \widehat{W}^{(0)} (x) \, = \,
    \exp \left[ -
    \int d \mu ( \widehat{P}_J ) \left(
    \frac{\del}{\del a_{J}^{(0)}} \otimes \frac{\del}{\del a_{-J}^{(0)}}
    \right) e^{ - i \widehat{P}_J \cdot ( x- y) }
    \right]_{y \rightarrow x}
    \label{4-37}
\eeq
where the limit $y \rightarrow x$ is taken so that the time ordering
$x^0 > y^0$ is preserved.
The contraction operator can thus be expressed as
\beq
    \widehat{W}^{(0)} (x) \, = \,
    \exp \left[ -
    \int \frac{d^4  \widehat{P}_J }{(2 \pi )^4} \frac{i}{ \widehat{p}_J^2 - m^2 }
    \left(
    \frac{\del}{\del a_{J}^{(0)}} \otimes \frac{\del}{\del a_{-J}^{(0)}}
    \right) e^{ - i \widehat{P}_J \cdot ( x- y) }
    \right]_{y \rightarrow x}  .
    \label{4-38}
\eeq
In deriving the above, we use the well-known identity
\beq
    \int d \mu ( q ) \left[
    \th ( x^0 - y^0 ) e^{-i q (x - y) } + \th ( y^0 - x^0 ) e^{i q (x - y)}
    \right]
    \, = \,
    \int \frac{ d^4 q} {(2 \pi )^4} \,
    \frac{i}{ q^2 - m^2 + i \ep }
    \, e^{- i q \cdot ( x - y ) }
    \label{4-39}
\eeq
where $q^\mu$ is an off-shell
four-momentum with mass $m$ and $\ep$ is a positive infinitesimal.

\noindent
\underline{An S-matrix functional for the UHV tree amplitudes}

We now apply the Wick-like contraction operator (\ref{4-38}) to the
generating functional (\ref{4-23}) for the UHV vertices:
\beq
    \F_{\rm UHV}  \left[ a^{(h)c}, a^{(0)} \right]
    ~ = ~
    \widehat{W}^{(0)} (x) \, \F_{\rm UHV}^{\rm (vertex)}  \left[ a^{(h)c}, a^{(0)} \right]
    \, .
    \label{4-40}
\eeq
By construction, we then find that
this functional serves as an S-matrix functional for the UHV tree amplitudes.
Explicitly, the UHV amplitudes in the $x$-space representation are generated as
\beqar
    &&
    \left.
    \frac{\del}{\del a_{1}^{(0)}} \otimes
    \frac{\del}{\del a_{2}^{(+)}} \otimes
    \frac{\del}{\del a_{3}^{(+)}} \otimes \cdots \otimes
    \frac{\del}{\del a_{n-1}^{(+)}} \otimes
    \frac{\del}{\del a_{n}^{(0)}}
    \, \F_{\rm UHV} \left[ a^{( \pm )c} , a^{(0)} \right]
    \right|_{a^{( \pm )c} = a^{(0)} = 0}
    \nonumber \\
    &=&
    \A_{\rm UHV}^{( \bar{\phi}_{1} g_{2}^{+} \cdots g_{n-1}^{+}  \phi_{n} )} (x)
    ~ \equiv ~
    \A_{\rm UHV}^{( \bar{\phi}_1  \phi_n )} (x) \, ,
    \label{4-41}
\eeqar
\beqar
    \A_{\rm UHV}^{( \bar{\phi}_1 \phi_n )} (x)
    & = &
    \int d \mu (\widehat{p}_1 )
    \prod_{i=2}^{n-1}  d \mu ( p_i )
    d \mu (\widehat{p}_n )
    \, \A_{\rm UHV}^{( \bar{\phi}_1 \phi_n )} ( u , \bu ) \, ,
    \label{4-42}\\
    \A_{\rm UHV}^{( \bar{\phi}_1 \phi_n )} ( u , \bu )
    &=&
    - i g^{n-2}
    \, (2 \pi)^4 \del^{(4)} \left( \widehat{p}_1 +
    \sum_{i=2}^{n-1} p_i + \widehat{p}_n \right) \,
    \widehat{A}_{\rm UHV}^{(  \phi_n \bar{\phi}_1 )} (u)
    \, ,
    \label{4-43} \\
    \widehat{A}_{\rm UHV}^{( \bar{\phi}_1  \phi_n )} (u)
    & = &
    \Tr ( t^{c_{2}} t^{c_{3}} \cdots t^{c_{n-1}}) \,
    \widehat{C}_{\rm UHV}^{(  \bar{\phi}_1 \phi_n  )} (u)
    \, ,
    \label{4-44} \\
    \widehat{C}_{\rm UHV}^{( \bar{\phi}_1  \phi_n )} (u)
    & = &
    \frac{ m^2 \, (n 1)}{ (12)(23) \cdots (n-1 \, n) (n 1) }
    \widehat{(n 1)}
    \, = \,
    A_{\rm UHV (0)}^{( \bar{\phi}_{1} g_{2}^{+} \cdots g_{n-1}^{+}  \phi_{n} ) }
    \label{4-45}
\eeqar
where we use the notation (\ref{4-32}) in the last equation.
Notice that here we are considering  $\widehat{(n 1)}$ and
$A_{\rm UHV (0)}^{( \bar{\phi}_{1} g_{2}^{+} \cdots g_{n-1}^{+}  \phi_{n})}$
in the $x$-space representation. Thus all the massive propagators
in (\ref{4-33}) and (\ref{4-44}) should be replaced by
\beq
    \frac{i}{\widehat{P}_J^2 - m^2} ~ \longrightarrow ~
    \frac{ d^4 \widehat{P}_J}{(4 \pi)^4} \frac{1}{\widehat{P}_J^2 - m^2}
    \label{4-46}
\eeq
or simply by $-i d \mu (\widehat{P}_J)$ in the expression (\ref{4-45}).
The extra factor $-i$ here comes from the definition of
the UHV vertex (\ref{4-25}) in terms of the generating functional (\ref{4-23}).
Owing to the saturation of Grassmann variables, there arise
no loop amplitudes made of the massive propagators for the UHV amplitudes\footnote{
The massive loop structure can, however, enter in purely gluonic part of
the UHV amplitudes. For example, we can easily consider massive one-loop
subamplitudes for all-plus gluon legs; these are relevant to the so-called
one-loop all-plus amplitudes in QCD. To incorporate these into the UHV
amplitudes, we need to introduce massless propagators
and an NUHV vertex. Thus such quantum effects do not arise as long as we
use the UHV S-matrix functional (\ref{4-40}). If we modify
the S-matrix to include these ingredients as we shall do in the next section,
however, the massive loop effects do arise in purely
gluonic part of the UHV amplitudes.}.
Notice also that the UHV amplitudes preserve
the single-trace structure due to the $U(1)$ color degrees of freedom
we assign to the massive scalars.

Since the UHV S-matrix functional (\ref{4-40}) leads to
Kiermaier's expression for the UHV tree amplitudes, the above
formulation shows nothing but a systematic
functional derivation of the massive CSW rules within the holonomy formalism,
at least for the UHV amplitudes.
We shall consider generalization to non-UHV cases in the next section.

Lastly, we comment that in our formalism the mass effect does not
break the supersymmetry. As analyzed in this section, the full
$\N = 4$ supersymmetry is crucial to derive the massive CSW rules.
The form of the UHV vertex (\ref{4-14}) suggests
breaking of supersymmetry from $\N = 4$ down to $\N = 2$ if we
treat the mass square in an isolated fashion.
We have tried to implement such an interpretation, {\it e.g.}, by
considering a peculiar off-shell continuation of the Nair measure or
by taking a different definition for supersymmetric massive operators,
but any attempts did not work well.
We thus come to realize that the holonomy formalism, or essentially
Nair's prescription for superamplitudes, can naturally be continued to
a massive system without breaking the extended $\N=4$ supersymmetry.

\section{An S-matrix functional for non-UHV tree amplitudes}

As in the non-MHV amplitudes of gluons,
the $n$-point non-UHV massive scalar amplitudes can be categorized as
N$^k$UHV amplitudes in terms of the number of negative-helicity
gluons, $k= 1,2, \cdots , n-3$.
The categorization is entirely gluonic since it dose not change
the physical information of massive scalars but that of gluons.
This means that such a categorization can be carried out
by the original CSW rules or the purely gluonic MHV rules.
Thus an S-matrix functional for massive scalar amplitudes in general can be
constructed as
\beqar
    \F^{(B)} \left[ a^{(h)c}, a^{(0)} \right] & = &
    \F_{\rm UHV}  \left[ a^{(h)c}, a^{(0)} \right]
    \,
    \F^{(A)}  \left[ a^{(h)c} \right]
    \nonumber \\
    &=&
    \widehat{W}^{(0)} (x) \,
    \widehat{W}^{(A)} (x) \,
    \F_{\rm UHV}^{\rm (vertex)}  \left[ a^{(h)c}, a^{(0)} \right]
    \,
    \F_{\rm MHV}  \left[ a^{(h)c} \right]
    \label{5-1}
\eeqar
where $\widehat{W}^{(0)} (x)$, $\widehat{W}^{(A)} (x)$,
$\F_{\rm UHV}^{\rm (vertex)}  \left[ a^{(h)c}, a^{(0)} \right]$ and
$\F_{\rm MHV}  \left[ a^{(h)c} \right]$ are given by
(\ref{4-37}), (\ref{2-34}), (\ref{4-23}) and (\ref{2-33}), respectively.

It is tempting to construct the S-matrix functional without use
of $\F^{(A)}  \left[ a^{(h)c} \right]$ since, as discussed in (\ref{3-6})-(\ref{3-8}),
the massive holonomy operator might include the purely gluonic
holonomy structure. However, once we fix the physical configuration space
$\C^{(B)}$ for the massive scalar system, such an inclusion becomes physically
difficult. For example, we need to separately define the braid traces of gluonic
and massive part of the operators.
There may be a way to circumvent these problems mathematically but
it seems too artificial and so far
we have not found any suitable methods
that would lead to a definition better than (\ref{5-1}).

\noindent
\underline{The NUHV tree amplitudes}

In what follows, we consider the next-to-UHV (NUHV) tree amplitudes, {\it i.e.},
the simplest non-UHV amplitudes that contain a pair of massive scalars,
one negative-helicity gluon, and an arbitrary number of positive-helicity gluons.
Using the S-matrix functional (\ref{5-1}), we can straightforwardly
calculate the holomoprhic NUHV tree amplitudes
$\widehat{A}_{\rm NUHV(0)}^{( \bar{\phi}_1  g_a^-  \phi_n )} (u)$,
the counterpart of $\widehat{A}_{\rm UHV}^{( \bar{\phi}_1  \phi_n )} (u)$
for the NUHV tree amplitudes, as
\beqar
    \widehat{A}_{\rm NUHV(0)}^{( \bar{\phi}_1  g_a^-  \phi_n )} (u)
    & = &
    \left.
    \sum_{i = 2}^{n-1} \sum_{r = 1}^{n-3}
    \widehat{A}^{( (i+r+1)_{+} \cdots \phi_n \bar{\phi}_1
    \cdots (i-1)_{+} l_{+} )}_{\rm UHV(0)} (u)
    \, \frac{1}{q_{i \, i+r}^2} \,
    \widehat{A}^{( (-l)_{-} \, (i)_{+} \cdots a_{-} \cdots (i+r)_{+} )}_{\rm MHV(0)} (u)
    \right|_{u_l = u_{i \, i+r}}
    \nonumber \\
    &=&
    \sum_{i = 2}^{n-1}
    \sum_{r = 1}^{n - 3 }
    \sum_{ \si \in \S_{r+1}} \!\!\!
    \Tr ( t^{\si_{i}} \cdots t^{\si_{i+r}}  \,
    t^{i+r+1} \cdots t^{i-1} )
    \nonumber \\
    &&
    \left.
    \widehat{C}^{( (i+r+1)_{+} \cdots \phi_n \bar{\phi}_1 \cdots (i-1)_{+} l_{+} )}_{\rm UHV(0)} (u)
    \, \frac{1}{q_{\si_{i} \si_{i+r}}^2} \,
    \widehat{C}^{( (-l)_{-} \, (i)_{+} \cdots a_{-} \cdots (i+r)_{+} )}_{\rm MHV(0)} (u; \si)
    \right|_{u_l = u_{\si_{i}  \si_{i+r} }}
    \label{5-2}
\eeqar
where, as usual, we consider the numbering indices in modulo $n$.
The $\widehat{C}$'s are therefore written as
\beqar
    \widehat{C}^{( (i+r+1)_{+} \cdots \phi_n \bar{\phi}_1
    \cdots (i-1)_{+} l_{+} )}_{\rm UHV(0)} (u)
    &=&
    \frac{m^2 (n1)  \widehat{(n1)}}{
    (i+r+1 ~ i+r+2) \cdots (i-1 ~ l)(l \, i+r+1 )
    } \, ,
    \label{5-3}\\
    \widehat{C}^{( (-l)_{-} \, (i)_{+} \cdots a_{-} \cdots (i+r)_{+} )}_{\rm MHV(0)} (u; \si)
    &=&
    \frac{(l a)^4 }{
    (l \, \si_i )( \si_i \, \si_{i+1} ) (\si_{i+1} \, \si_{i+ 2} ) \cdots (\si_{i+r} \, l)
    }
    \label{5-4}
\eeqar
where the off-shell momentum transfer and an associated
spinor momentum are defined as
\beqar
    q_{i \, i+r}^{A \Ad}
    &=&
    p_{i}^{A \Ad} + p_{i+1}^{A \Ad} + \cdots + p_{i+r}^{A \Ad}
    \, \equiv \, p_{i \, i+r}^{A \Ad} + w \eta_{i \, i+r}^{A \Ad}
    \, ,
    \label{5-5} \\
    p_{i \, i+r}^{A \Ad} & = & u_{i \, i+r}^A \bu_{i \, i+r}^\Ad
    \, \equiv \, u_{l}^A \bu_{l}^\Ad
    \, .
    \label{5-6}
\eeqar
Here $w$ is a real number and $\eta_{i \, i+r}^{A \Ad}$ is a reference null-vector.
Permutation of the numbering indices for gluons is represented by
\beq
    \si=\left(%
    \begin{array}{l}
      i ~~\, i+1 ~ \cdots ~ i+r  \\
      \si_{i} ~~ \si_{i+1} ~ \cdots ~ \si_{i+r} \\
    \end{array}%
    \right)  .
    \label{5-7}
\eeq
Accordingly, the indices of the momentum transfer (\ref{5-5}) are
labeled by $\si_i$'s under the permutation $\si \in \S_{r+1}$.
In the expression (\ref{5-2}), we denote $u_{i \, i+r } \equiv u_l$, or
$u_{\si_{i} \, \si_{i+r} } \equiv u_l$ under the permutation, for simplicity.
Also the $SU(N)$ generators $t^{c_{{\si}_i}}$'s are abbreviated by $t^{\si_i}$'s.

Notice that the structure of the NUHV tree amplitudes is
the same as the NMHV tree amplitudes of gluons
except that in the former cases
one of the MHV vertices is replaced by the UHV vertex.
Consequently, there appear no sums over permutations over the
numbering indices involving the UHV vertex.

\noindent
\underline{Examples of the NUHV tree amplitudes and comparison with BS expressions}

By construction, there are no 3-point NUHV tree amplitudes. Non-vanishing
NUHV tree amplitudes start from $n \ge 4$.
In what follows we consider first few examples of these.
For $n=4$, we can write down the NUHV tree amplitude as
\beqar
    \widehat{A}_{\rm NUHV(0)}^{( \bar{\phi}_1 g_2^+ g_3^-  \phi_4 )} (u)
    &=&
    \left.
    \widehat{A}^{( \phi_4 \bar{\phi}_1 l_{+} )}_{\rm UHV(0)} (u)
    \, \frac{1}{q_{2 3}^2} \,
    \widehat{A}^{( (-l)_{-} \, 2_{+}  3_{-} )}_{\rm MHV(0)} (u)
    \right|_{u_l = u_{2 3}}
    \nonumber \\
    &=&
    \sum_{ \si \in \S_{2}}
    \Tr ( t^{1} t^{\si_{2}} t^{\si_{3}} t^{4}  )
    \,
    \left.
    \widehat{C}^{( \phi_4 \bar{\phi}_1 l_{+} )}_{\rm UHV(0)} (u)
    \, \frac{1}{q_{\si_2 \si_3}^2} \,
    \widehat{C}^{( (-l)_{-} \, 2_{+}  3_{-} )}_{\rm MHV(0)} (u; \si)
    \right|_{u_l = u_{\si_2  \si_3 }}
    \nonumber \\
    &=&
    \left.
    \Tr ( t^{2} t^{3}  )
    \frac{m^2 (41)}{(1 l) (l 4) } \frac{1}{q_{23}^2}
    \left(
    \frac{(l3)^4}{(l2)(23)(3l)}
    +     \frac{(l3)^4}{(l3)(32)(2l)}
    \right)
    \right|_{u_l = u_{23}}
    \label{5-8}
\eeqar
where we use the fact that the color factor of the massive
scalars is assigned to the $U(1)$ direction of the
$U(N) = SU(N) \times U(1)$ gauge group.
Notice that the invariance under permutations of gluon legs is
explicit in the above expression.
In the literature this invariance is implicit and
the amplitudes are usually given in a form of the first line in (\ref{5-8}).
Taking account of this fact, we find that the above expression
agrees with the previously known result by Boels and Schwinn
\cite{Boels:2007pj} with a certain choice of
the reference spinor for $\bar{\phi}_1$.

To be more precise, we can fix the reference spinor
such that the on-shell partner $p_{1}^{\mu}$ of $\widehat{p}_{1}^{\mu}$
is proportional to $p_{2}^{\mu}$ or
$u_1^A \parallel u_2^A$ where
$p_{2}^{\mu}$ denotes the gluon four-momentum with the numbering index 2.
By doing so, we can easily see that the
4-point NUHV tree amplitudes in equation (2.10) of \cite{Boels:2007pj},
the Boels-Schwinn (BS) expression, reduces to the above expression (\ref{5-8})
since those terms proportional to $( u_1 u_2 ) = (12)$ vanishes
in the BS expression.
Notice that the BS expression is stripped of
color factors and permutation invariance under gluon transpositions.
To compare the BS expression with our result (\ref{5-8}), note also that
the BS expression has a different helicity configuration from (\ref{5-8});
the negative helicity is assigned to $g_3$, not to $g_2$, in there.

The reference spinor of the above choice can be specified by the one satisfying
$\eta_{1}^{\mu} = \frac{2 c (p_2 \cdot \eta_1 )}{m^2}
( \widehat{p}_{1}^{\mu} - c p_{2}^{\mu} )$ where $c$ is a constant.
In the BS expression, the reference spinors for the massive scalars are
set to identical, contrary to our choice in (\ref{4-20})-(\ref{4-22}).
Thus one can effectively reduce the BS expression to a form
which is more compact than (\ref{5-8}) by choosing a suitable
reference spinor. In fact, such a choice was made in
equation (2.11) of \cite{Boels:2007pj}.
What we have shown here is that an alternative choice of
the reference spinor leads to a different reduction
of the BS expression where only the terms that
involve the massless propagators survive.
This reflects our basic relations in (\ref{4-18}) and (\ref{4-19}),
{\it i.e.}, the vanishing of the NUHV and UHV$^2$ vertices,
in contrast to the BS relations in  (\ref{4-15}) and (\ref{4-16}).

For $n=5$ we can similarly compute the NUHV tree amplitudes as follows:
\beqar
    &&
    \!\!\!\!\!\!\!\!\!\!
    \widehat{A}_{\rm NUHV(0)}^{( \bar{\phi}_1 g_2^+ g_3^+ g_4^-  \phi_5 )} (u)
    \nonumber \\
    &=&
    \left.
    \widehat{A}^{( \phi_5 \bar{\phi}_1 l_{+} )}_{\rm UHV(0)} (u)
    \, \frac{1}{q_{2 4}^2} \,
    \widehat{A}^{( (-l)_{-} \, 2_{+}  3_{+} 4_{-} )}_{\rm MHV(0)} (u)
    \right|_{u_l = u_{24}}
    +
    \left.
    \widehat{A}^{( \phi_5 \bar{\phi}_1 2_+ l_{+} )}_{\rm UHV(0)} (u)
    \, \frac{1}{q_{3 4}^2} \,
    \widehat{A}^{( (-l)_{-} \,  3_{+} 4_{-} )}_{\rm MHV(0)} (u)
    \right|_{u_l = u_{34}} \, ,
    \nonumber \\
    \label{5-9}\\
    &&
    \!\!\!\!\!\!\!\!\!\!
    \widehat{A}_{\rm NUHV(0)}^{( \bar{\phi}_1 g_2^+ g_3^- g_4^+  \phi_5 )} (u)
    \nonumber \\
    &=&
    \left.
    \widehat{A}^{( 4_+ \phi_5 \bar{\phi}_1 l_{+} )}_{\rm UHV(0)} (u)
    \, \frac{1}{q_{23}^2} \,
    \widehat{A}^{( (-l)_{-} \, 2_{+} 3_{-} )}_{\rm MHV(0)} (u)
    \right|_{u_l = u_{23}}
    +
    \left.
    \widehat{A}^{( \phi_5 \bar{\phi}_1 l_{+} )}_{\rm UHV(0)} (u)
    \, \frac{1}{q_{24}^2} \,
    \widehat{A}^{( (-l)_{-} \, 2_+ 3_{-} 4_{+} )}_{\rm MHV(0)} (u)
    \right|_{u_l = u_{24}}
    \nonumber \\
    &&
    \, +  \,
    \left.
    \widehat{A}^{( \phi_5 \bar{\phi}_1 2_+ l_{+} )}_{\rm UHV(0)} (u)
    \, \frac{1}{q_{3 4}^2} \,
    \widehat{A}^{( (-l)_{-} \,  3_{-} 4_{+} )}_{\rm MHV(0)} (u)
    \right|_{u_l = u_{34}}  \, .
    \label{5-10}
\eeqar
The rest of the 5-point NUHV amplitudes,
$\widehat{A}_{\rm NUHV(0)}^{( \bar{\phi}_1 g_2^- g_3^+ g_4^+  \phi_5 )} (u)$,
can also be obtained from a symmetry argument.
The expression (\ref{5-9}) is in accord with the previously known result
\cite{Boels:2008ef} with a certain choice of the reference spinor for $\bar{\phi}_1$
in the same sense that we have argued in the case of 4-point amplitudes.

To be more concrete, one can reduce the Boels-Schwinn expression
for the 5-point NUHV tree amplitude, the one given
in equation (3.10) of \cite{Boels:2008ef},
to the form of (\ref{5-9}) by choosing the on-shell partner
of the massive scalar $\bar{\phi}_1$ to be proportional to
$p_2^\mu$ in the BS expression.
As in the case of the 4-point amplitude, such a choice
removes all the contributions that do not contain
massless propagators transferred by virtual gluons,
and leads to the expression (\ref{5-9}) once
proper color structure and permutation invariance
under gluon exchanges are imposed.

As studied in \cite{Boels:2007pj,Boels:2008ef},
the BS expressions for the 4-point and the 5-point
NUHV tree amplitudes numerically agree with other
set of NUHV tree amplitudes \cite{Badger:2005zh,Forde:2005ue}
obtained by BCFW-type recursion methods.
It is interesting to find that the above analyses
show the connection between our
formalism and the BS expressions at least for
the cases of (\ref{5-8}) and (\ref{5-9}).
For the other NUHV amplitudes, there exist no BS-type
amplitudes in the literature. In this sense, generalization
of the above analyses to $n > 5$ is not clear at this stage,
however, we shall observe the appropriateness of our formalism
for arbitrary $n$ in the next subsection from a different perspective.

For further understanding of our formalism we now present
the 6-point NUHV tree amplitudes below:
\beqar
    &&
    \!\!\!\!\!\!\!\!\!\!
    \widehat{A}_{\rm NUHV(0)}^{( \bar{\phi}_1 g_2^+ g_3^+ g_4^+ g_5^- \phi_6 )} (u)
    \nonumber \\
    &=&
    \left.
    \widehat{A}^{( \phi_6 \bar{\phi}_1 l_{+} )}_{\rm UHV(0)} (u)
    \, \frac{1}{q_{25}^2} \,
    \widehat{A}^{( (-l)_{-} \, 2_{+} 3_{+} 4_{+} 5_{-} )}_{\rm MHV(0)} (u)
    \right|_{u_l = u_{25}}
    +
    \left.
    \widehat{A}^{( \phi_6 \bar{\phi}_1 2_+ l_{+} )}_{\rm UHV(0)} (u)
    \, \frac{1}{q_{35}^2} \,
    \widehat{A}^{( (-l)_{-} \, 3_{+} 4_{+} 5_{-} )}_{\rm MHV(0)} (u)
    \right|_{u_l = u_{35}}
    \nonumber \\
    &&
    \, +  \,
    \left.
    \widehat{A}^{( \phi_6 \bar{\phi}_1 2_{+} 3_{+} l_{+} )}_{\rm UHV(0)} (u)
    \, \frac{1}{q_{45}^2} \,
    \widehat{A}^{( (-l)_{-} \,  4_{+} 5_{-} )}_{\rm MHV(0)} (u)
    \right|_{u_l = u_{45}}  \, ,
    \label{5-11}\\
    &&
    \!\!\!\!\!\!\!\!\!\!
    \widehat{A}_{\rm NUHV(0)}^{( \bar{\phi}_1 g_2^+ g_3^+ g_4^- g_5^+ \phi_6 )} (u)
    \nonumber \\
    &=&
    \left.
    \widehat{A}^{( \phi_6 \bar{\phi}_1 l_{+} )}_{\rm UHV(0)} (u)
    \, \frac{1}{q_{25}^2} \,
    \widehat{A}^{( (-l)_{-} \, 2_{+} 3_{+} 4_{-} 5_{+} )}_{\rm MHV(0)} (u)
    \right|_{u_l = u_{25}}
    +
    \left.
    \widehat{A}^{( 5_+ \phi_6 \bar{\phi}_1 l_{+} )}_{\rm UHV(0)} (u)
    \, \frac{1}{q_{24}^2} \,
    \widehat{A}^{( (-l)_{-} \, 2_{+} 3_{+} 4_{-} )}_{\rm MHV(0)} (u)
    \right|_{u_l = u_{24}}
    \nonumber \\
    &&
    \, + \,
    \left.
    \widehat{A}^{( \phi_6 \bar{\phi}_1 2_{+} l_{+} )}_{\rm UHV(0)} (u)
    \, \frac{1}{q_{35}^2} \,
    \widehat{A}^{( (-l)_{-} \, 3_{+} 4_{-} 5_{+} )}_{\rm MHV(0)} (u)
    \right|_{u_l = u_{35}}
    +
    \left.
    \widehat{A}^{( 5_+ \phi_6 \bar{\phi}_1 2_+ l_{+} )}_{\rm UHV(0)} (u)
    \, \frac{1}{q_{34}^2} \,
    \widehat{A}^{( (-l)_{-} \, 3_{+} 4_{-} )}_{\rm MHV(0)} (u)
    \right|_{u_l = u_{34}}
    \nonumber \\
    &&
    \, +  \,
    \left.
    \widehat{A}^{( \phi_6 \bar{\phi}_1 2_{+} 3_{+} l_{+} )}_{\rm UHV(0)} (u)
    \, \frac{1}{q_{45}^2} \,
    \widehat{A}^{( (-l)_{-} \,  4_{-} 5_{+} )}_{\rm MHV(0)} (u)
    \right|_{u_l = u_{45}} \, .
    \label{5-12}
\eeqar

As far as the author notices, there exist no explicit
expressions for the NUHV tree amplitudes beyond $n = 5$
except the recent calculation of the so-called
one-minus amplitudes in QCD at one-loop level,
{\it i.e.}, one-loop amplitudes of one negative-helicity gluon
and an arbitrary number of positive-helicity gluons, with the
internal lines being massive scalar propagators.
The calculation is carried out by Elvang, Freedman and Kiermaier (EFK) in
\cite{Elvang:2011ub}, applying the
massive CSW rules for the one-loop amplitudes in QCD.
The resultant one-minus one-loop amplitudes can easily
be rendered into a form of the NUHV tree amplitudes and
we can compare them with our general expression (\ref{5-2}).
For this purpose, we now briefly review the EFK results
for the one-loop one-minus amplitudes and their applications
to the NUHV tree amplitudes for arbitrary $n$.

\noindent
\underline{Comparison with the EFK representation for the NUHV tree amplitudes}

The EFK calculation for the one-loop one-minus amplitudes
is given in a form of integrand which corresponds to the holomorphic amplitudes
$\widehat{A}_{\rm NUHV(0)}^{( {\bar \phi}_{1}
g_2^- g_3^+ \cdots g_{n-1}^{+} \phi_n )} (u)$.
One should also notice that the dependence on the reference spinors of
(internal) massive scalars are kept explicit in the EFK calculation, {\it i.e.},
the reference spinors are kept unspecified throughout the calculation.
This is for the purpose of verifying the absence of spurious poles
by use of the $\eta$-independence of the amplitudes; see \cite{Elvang:2011ub}
for details.

\begin{figure} [htbp]
\begin{center}
\includegraphics[width=140mm]{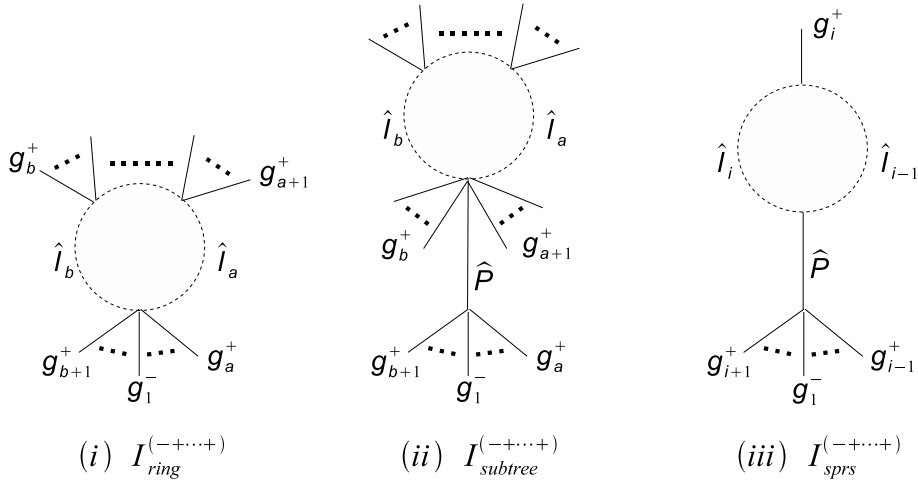}
\caption{Diagrams contributing to the one-loop one-minus amplitudes of $n$ gluons}
\label{fighol0501}
\end{center}
\end{figure}

Following the notation in \cite{Elvang:2011ub}, the EFK result for the
$n$-point one-loop one-minus amplitudes/integrands can be summarized as
\beq
    I^{(1_- 2_+ 3_+ \cdots n_+ )} \, = \,
    I^{(-+ \cdots +)}_{ring} + I^{(-+ \cdots +)}_{subtree} + I^{(-+ \cdots +)}_{sprs} \, ,
    \label{5-13}
\eeq
where the terms in the right-hand side respectively correspond to the
diagrams in Figure \ref{fighol0501} and are explicitly given by
\beqar
    I^{(-+ \cdots +)}_{ring}
    &=&
    \sum_{ a < b} \frac{ - 2 N_p m^2 (1 l_a )^2 (1 l_b )^2 (a \, a+1) (b \, b+1)}{
    (12)(23) \cdots (n1) \, (l_a l_b ) ( a l_a ) (l_a \, a+1 )(b l_b ) (l_b \, b+1)
    }
    \frac{i}{ \widehat{l}_a^2 - m^2 } \frac{i}{ \widehat{l}_b^2 - m^2 }
    \nonumber \\
    &&
    ~~~ \times
    \left( l_a \left|
    \prod_{j = a+1}^{b-1} \left[
    1 - \frac{  i m^2 |l_j ) ( j \, j+1 ) ( l_j | }
    {(\widehat{l}_j^2 - m^2) ( j \, l_j ) (l_j \, j+1 ) }
    \right]
    \right| l_b \right)  \, ,
    \label{5-14}\\
    I^{(-+ \cdots +)}_{subtree}
    &=&
    \sum_{ 2 \le b-a \le n-2}
    \frac{(1 P)^4}{
    (P \, b+1)(b+1 \, b+2) \cdots (a-1 \, a) (a P)}
    \nonumber \\
    &&
    ~~~~~~~~~~~~~~ \times
    \, \frac{i}{\widehat{P}^2} ~
    I^{(++ \cdots +)}_{CSW} (a+1, \cdots, b, P) \, ,
    \label{5-15}\\
    I^{(-+ \cdots +)}_{sprs}
    &=&
    \sum_{i=2}^{n}
    \frac{ - 2N_p m^2 }{ (12)(23) \cdots (n1) }
    \frac{\sqrt{-1}}{ \widehat{l}_i^2 - m^2 } \frac{\sqrt{-1}}{ \widehat{l}_{i-1}^{2} - m^2 }
    \frac{(1 i)^2 }{ (i l_i )}
    \nonumber \\
    &&
    ~~~ \times
    \left[
    \frac{(1 l_{i-1}) (1 l_i )}{(i l_i )}
    -  \frac{(1 \, i-1) (1 l_i )}{(i-1 \,  i )}
    +   \frac{(1 \, i+1 ) (1 l_{i-1} )}{(i \, i+1 )}
    \right] \, .
    \label{5-16}
\eeqar
In (\ref{5-15}) the one-loop all-plus integrand
$I^{(++ \cdots +)}_{CSW} (a+1, \cdots, b, P)$ is defined
as a trace over the UHV tree amplitudes:
\beqar
    I^{(++ \cdots +)}_{CSW} (a+1, \cdots, b, P) & = &
    \frac{2 N_p}{(a+1 \, a+2)(a+2 \, a+3)\cdots (b P)(P \, a+1)}
    \nonumber \\
    &&
    \times
    \, \Tr^\prime \prod_{j=a+1}^{P} \left[
    1 - \frac{ i m^2 | l_j ) (j \, j+ 1 ) (l_j |}{ ( \widehat{l}_j^2 - m^2 )
    (j \, l_j ) ( l_j \, j+1 ) }
    \right]
    \label{5-17}
\eeqar
where the trace is taken over the two-component spinors and $\Tr^\prime$ is defined as
$\Tr^\prime X \equiv \Tr X - \Tr {\bf 1}$ to subtract the identity factor $\Tr {\bf 1} =2$
from (\ref{5-17}).
In the expressions (\ref{5-14})-(\ref{5-17}), the overall
prefactor $N_p$ is introduced in order to make the pure gluonic amplitudes to
those of QCD including (massless) fermion contributions in the loop.
Neglecting the fermion contributions, we can set $N_p = 1$.

To obtain the NUHV tree amplitudes from these EFK results of
the one-loop one-minus QCD amplitudes, we should bear in mind
the following two things.
\begin{enumerate}
  \item One is the fact that in a one-loop (MHV) diagram we can make at least one
leg on each side of the diagram be collinear to each other. This is due
to the freedom we have in the choice of the reference spinors involving the
loop propagators. This is also a useful lesson we have learned from
the one-loop calculations of $\N = 4$ super Yang-Mills theory in
the holonomy formalism, see \cite{Abe:2011af} for details.
  \item The other thing is, as mentioned earlier, that the reference spinors
involved in the EFK results (\ref{5-13})-(\ref{5-17}) are not specified.
This has been convenient to study the freedom from spurious poles
in the one-loop amplitudes \cite{Elvang:2011ub}.
However, for our purposes, {\it i.e.} to
compare our formulation of the NUHV tree amplitudes (\ref{5-2}) to
the EFK results (\ref{5-13})-(\ref{5-17}), we no longer need to
keep the reference spinors arbitrary. We can fix them in a suitable way
before reducing the EFK results to the NUHV tree amplitudes.
\end{enumerate}

From the first condition, we can easily find that the ring integrand
$I^{(-+ \cdots +)}_{ring}$ vanishes upon the choice of
$u_a \parallel u_{a+1}$, {\it i.e.}, $(a \, a+1) = 0$.
The first condition also implies that diagram $(iii)$ in Figure \ref{fighol0501}
can be treated as a tadpole-like diagram. This means that
the integrand $I^{(-+ \cdots +)}_{sprs}$ can be considered as
a UHV-MHV type amplitude upon the reduction to the NUHV tree amplitude
by cutting the massive loop apart. The UHV part of the tree amplitude then
has the minimum three legs composed of two massive scalars and one
internal virtual gluon (but no pure gluons).
Similarly, we can reduce the subtree integrand
$I^{(-+ \cdots +)}_{subtree}$ to UHV-MHV type amplitudes upon
cutting the massive loop apart. The UHV part of the reduced NUHV tree amplitudes
have more than three legs (including an arbitrary number of gluons).
These analyses show that the EFK representation for the NUHV tree amplitudes is given
in terms of UHV and MHV vertices connected by massive scalar
propagators. This description agrees with our construction
of the NUHV tree amplitudes (\ref{5-2}). In fact, installing
information of color factors and permutation invariance under
gluon transpositions, we find that the above EFK representation of the NUHV
tree amplitudes exactly agrees with our formulation (\ref{5-2}).

\noindent
\underline{Generalization to non-UHV tree amplitudes}

Field theoretically it is straightforward to obtain non-UHV tree amplitudes
out of the S-matrix functional for massive scalar amplitudes in (\ref{5-1}).
We simply apply a sequence of functional derivatives of interest to the
S-matrix functional and evaluate the derivatives as in (\ref{4-41}).
Because of the contraction operators the computation is entirely
based on the massless and massive CSW rules except that
we make use of vanishing non-UHV vertices (\ref{4-18}), (\ref{4-19})
in the massive part, rather than the original proposal (\ref{4-15}), (\ref{4-16}).
As in the cases of non-MHV gluon amplitudes, we can hence
obtain non-UHV massive scalar amplitudes
in terms of the UHV and the MHV vertices or what we previously call the
``UHV rules.''
An explicit form of the non-UHV amplitudes are as tedious and complicated
as that of the non-MHV amplitudes. From the S-matrix functional (\ref{5-1}), however,
we can easily obtain a succinct recursive expression for the non-UHV tree amplitudes:
\beqar
    &&
    \widehat{A}_{\rm N^k UHV(0)}^{
    ({\bar \phi}_1 g_{a_1}^{-} g_{a_2}^{-} \cdots g_{a_k}^{-} \phi_n )
    } (u)
    \nonumber \\
    & = &
    \left.
    \sum_{i = 2}^{n-1} \sum_{r = 1}^{n-3}
    \widehat{A}^{( (i+r+1)_{+} \cdots \phi_n \bar{\phi}_1
    \cdots (i-1)_{+} l_{+} )}_{\rm UHV(0)} (u)
    \, \frac{1}{q_{i \, i+r}^2} \,
    \widehat{A}^{( (-l)_{-} \, (i)_{+} \cdots a_{1-} \cdots
    a_{2-} \cdots \cdots a_{k-} \cdots (i+r)_{+} )}_{\rm N^{k-1}MHV(0)} (u)
    \right|_{u_l = u_{i \, i+r}}
    \label{5-18}
\eeqar
where $k= 1,2, \cdots , n-3$ and
the meanings of $q_{i \, i+r}$ and $u_l$ are the same as in (\ref{5-2}).

Alternatively, we can also express the N$^k$UHV massive scalar amplitudes as
\beqar
    &&
    \widehat{A}_{\rm N^k UHV(0)}^{
    ({\bar \phi}_1 g_{a_1}^{-} g_{a_2}^{-} \cdots g_{a_k}^{-} \phi_n )
    } (u)
    \nonumber \\
    & = &
    \left.
    \sum_{i = 2}^{n-1} \sum_{r = 1}^{n-3}
    \widehat{A}^{( (i+r+1)_{+} \cdots \phi_n \bar{\phi}_1
    \cdots (i-1)_{+} l_{+} )}_{\rm N^{k-1}UHV(0)} (u)
    \, \frac{1}{q_{i \, i+r}^2} \,
    \widehat{A}^{( (-l)_{-} \, (i)_{+} \cdots a_{k -} \cdots (i+r)_{+} )}_{\rm MHV(0)} (u)
    \right|_{u_l = u_{i \, i+r}}
    \label{5-19}
\eeqar
where we make the negative-helicity indices implicit in
$\widehat{A}^{( (i+r+1)_{+} \cdots \phi_n \bar{\phi}_1
 \cdots (i-1)_{+} l_{+} )}_{\rm N^{k-1}UHV(0)} (u)$.
The latter expression (\ref{5-19}) is recursive in terms of the
massive part of the massive amplitudes, while the former (\ref{5-18}) is
in terms of the gluonic part of them.
The gluonic N$^k$MHV tree amplitudes in (\ref{5-18}) can be constructed by
the original massless CSW rules, while the massive N$^k$UHV tree amplitudes
in (\ref{5-19}) is constructed only from the N$^{k-1}$UHV counterpart
in an inductive way.
In this sense the two expressions reveal the structure behind
the CSW rules and the BCFW recursion methods and indicate
their equivalence in the calculations of the massive scalar amplitudes at tree level.

Lastly we should comment on loop effects to the massive scalar amplitudes.
As noted earlier, massive loop effects arise in purely
gluonic part of the massive scalar amplitudes.
For example, we can incorporate one-loop
all-plus gluon configurations into the gluonic part of the UHV amplitudes.
There are also massless loop effects contributing to the
general gluon amplitudes.
Therefore, taking the expression (\ref{5-18}) for instance, we can
{\it in principle} calculate the one-loop massive scalar amplitudes as
\beqar
    &&
    \widehat{A}_{\rm N^k UHV(1)}^{
    ({\bar \phi}_1 g_{a_1}^{-} g_{a_2}^{-} \cdots g_{a_k}^{-} \phi_n )
    } (u)
    \nonumber \\
    & = &
    \sum_{i = 2}^{n-1} \sum_{r = 1}^{n-3}
    \left[
    \widehat{A}^{( (i+r+1)_{+} \cdots \phi_n \bar{\phi}_1
    \cdots (i-1)_{+} l_{+} )}_{\rm UHV(1)} (u)
    \, \frac{1}{q_{i \, i+r}^2} \,
    \widehat{A}^{( (-l)_{-} \, (i)_{+} \cdots a_{1-} \cdots
    a_{2-} \cdots \cdots a_{k-} \cdots (i+r)_{+} )}_{\rm N^{k-1}MHV(0)} (u)
    \right.
    \nonumber \\
    &&
    ~~~
    + \,
    \left.
    \widehat{A}^{( (i+r+1)_{+} \cdots \phi_n \bar{\phi}_1
    \cdots (i-1)_{+} l_{+} )}_{\rm UHV(0)} (u)
    \, \frac{1}{q_{i \, i+r}^2} \,
    \widehat{A}^{( (-l)_{-} \, (i)_{+} \cdots a_{1-} \cdots
    a_{2-} \cdots \cdots a_{k-} \cdots (i+r)_{+} )}_{\rm N^{k-1}MHV(1)} (u)
    \right]_{u_l = u_{i \, i+r}}
    \label{5-20}
\eeqar
From this expression we may obtain explicit forms of the one-loop
massive scalar amplitudes but that task is beyond the scope of
the present paper and we shall leave it to future works.

\section{Concluding remarks}

One of the main purposes of this paper is to investigate
whether an off-shell continuation of Nair's superamplitude method
can be applied to a massive model, particularly to amplitudes
of an arbitrary number of gluons and
a pair of massive scalars (or massive scalar amplitudes),
in a framework of the holonomy formalism.
In the present paper we have affirmed this proposition by
making a specific choice of the reference spinors for the
massive scalars, see (\ref{4-20}).
This allows us to obtain a functional derivation of the so-called
ultra helicity violating (UHV) vertices (\ref{4-14}) of the massive scalar
amplitudes, eventually leading us to the S-matrix functional (\ref{5-1})
for the massive scalar amplitudes in general.

An essential ingredient of the S-matrix functional is
given by the massive holonomy operator $\Theta_{R , \ga}^{(B)} (u)$
we have defined in section 3.
This operator is relevant to the generation of the massive
part of the massive scalar amplitudes. In particular,
together with the contraction operator $\widehat{W}^{(0)} (x)$
in (\ref{4-37}), this operator generates
the UHV tree amplitudes (\ref{4-41})-(\ref{4-45})
in a form that was previously reported by Kiermaier \cite{Kiermaier:2011cr},
using the so-called massive CSW rules of
Boels and Schwinn \cite{Boels:2007pj,Boels:2008ef}.

One of the interesting features in our formulation is that
a careful analysis of the braid trace in the construction of
$\Theta_{R , \ga}^{(B)} (u)$ leads to no apparent sums over
permutations of gluons in the color structure of the UHV tree amplitudes,
or the massive part of the massive scalar amplitudes in general.
The number of terms involving the UHV tree amplitudes, however, does not
drastically decrease from that of the MHV tree amplitudes of gluons.
This is due to the fact that
actions of the contraction operator $\widehat{W}^{(0)} (x)$
do not alter the gluon helicity configurations for the massive scalar
amplitudes. We have briefly presented a quantitative analysis of
this fact in (\ref{4-35}) and below.

The holonomy formalism is neither Lagrangian
nor Hamiltonian formalism so that we do/can not
introduce potentials for the incorporation of massive particles.
Information of mass is embedded into physical operators such
that helicity or polarization of the particles of interest is
in accord with the definition of the helicity operator (\ref{2-31}).
In practice, this can be carried out by considering
an off-shell continuation of Nair's superamplitude method.
In the present paper we strictly follow this idea, with
its concrete realization given in (\ref{4-6}).
Notice that our approach is philosophically different from
other approaches found in the literature.
For example, the massive CSW rules \cite{Boels:2007pj,Boels:2008ef}
are derived from the Lagrangian formalism \cite{Gorsky:2005sf}-\cite{Boels:2007qn}.
The earlier approaches \cite{Dixon:2004za}-\cite{Ferrario:2006np}, on the other hand,
focus more on the application of the CSW rules or
the BCFW recursion relations to massive models and its usage rather than
its derivation from first principles.
The holonomy formalism is therefore qualitatively new in the
studies of massive scalar amplitudes and
possibly leads to a completely new mass generation mechanism.
Obviously it is worth investigating how massive fermions and massive bosons
will be incorporated into the same framework.
We shall consider such extensions in a forthcoming paper.

Lastly, we would like to emphasize that in our formalism
mass effects do not invoke supersymmetry breaking.
The full $\N = 4$ supersymmetry has been crucial to derive
the correct form of the UHV vertex (\ref{4-14})
which forms a basic building block for the massive scalar amplitudes.
(Such a construction is referred to as the ``UHV rules'' in the above text.)
As explicitly shown in (\ref{4-23})-(\ref{4-26}),
the derivation is given in a conventional functional method.
Thus the holonomy formalism can naturally be continued to
a massive model without breaking the extended $\N =4$ supersymmetry.
It would be useful to take account of this fact in
the construction of more realistic massive models
in the framework of holonomy formalism or, more broadly, in the
four-dimensional spinor-helicity formalism.


\end{document}